\documentclass[aps,prx,preprint,onecolumn,citeautoscript,nofootinbib,eqsecnum]{revtex4-1}
\usepackage[T1]{fontenc}
\usepackage{feynmp}
\usepackage{comment}
\usepackage{subcaption}
\usepackage{amsmath}
\usepackage{amssymb}
\usepackage{bbm}
\usepackage{braket}
\usepackage{xcolor}
\usepackage{pifont}
\usepackage{slashed}
\usepackage[mathscr]{euscript}
\usepackage[shortlabels]{enumitem}
\usepackage{tikz-feynman}
\usepackage[papersize={8.5in,11in}]{geometry}
\usepackage{float}
\allowdisplaybreaks

\usepackage{graphicx}
\usepackage{subcaption}
\usepackage{multirow}
\usepackage[colorlinks=true]{hyperref}
\renewcommand{\approx}{\simeq}

\renewcommand{\Im}{\text{Im}}

\newcommand{\ce}{\mathcal{E}}

\definecolor{wrongultramarine}{rgb}{1,0.5,0}

\newcommand{\rd}{{\rm d}}
\newcommand{\sgn}{{\rm sgn\,}}

\newcommand{\beq}{\begin{equation}}
\newcommand{\eeq}{\end{equation}}
\def\bea{\begin{eqnarray}}
\def\eea{\end{eqnarray}}

\newcommand{\cK}{\mathcal{K}}
\newcommand{\cN}{\mathcal{N}}
\newcommand{\cD}{\mathcal{D}}

\hypersetup{
    bookmarks=true,         
    unicode=false,          
    pdftoolbar=true,        
    pdfmenubar=true,        
    pdffitwindow=false,     
    pdfstartview={FitH},    
    pdfsubject={},   
    pdfcreator={},   
    pdfproducer={}, 
    pdfkeywords={} {} {}, 
    pdfnewwindow=true,      
    colorlinks=true,       
    linkcolor=blue, 
    citecolor=blue,        
    filecolor=magenta,      
    urlcolor=blue           
}

\tikzset{
  mid arrow/.style={postaction={decorate,decoration={
        markings,
        mark=at position .575 with {\arrow[#1]{stealth}}
      }}},
  near arrow/.style={postaction={decorate,decoration={
        markings,
        mark=at position .275 with {\arrow[#1]{stealth}}
      }}},
   far arrow/.style={postaction={decorate,decoration={
        markings,
        mark=at position .800 with {\arrow[#1]{stealth}}
      }}},
}

\begin{document}
\title{Extrinsic phonon thermal Hall transport from Hall viscosity}

\author{Haoyu Guo}
\affiliation{Department of Physics, Harvard University, Cambridge MA 02138, USA}

\author{Subir Sachdev}
\affiliation{Department of Physics, Harvard University, Cambridge MA 02138, USA}

\date{\today}

\begin{abstract}
Motivated by recent experiments on the phonon contribution to the thermal Hall effect in the cuprates, we present an analysis of chiral phonon transport.
We assume the chiral behavior arises from a non-zero phonon Hall vicosity, which is likely induced by the coupling to electrons.
Phonons with a non-zero phonon Hall viscosity have an intrinsic thermal Hall conductivity, but Chen {\it et al.\/} (Phys. Rev. Lett. {\bf 124}, 167601 (2020)) have argued that a significantly larger thermal Hall conductivity can arise from an extrinsic contribution which is inversely proportional to the density of impurities. We solve the Boltzmann equation for phonon transport and compute the temperature ($T$) dependence of the thermal Hall conductivity originating from skew scattering off point-like impurities. We find that the dominant source for thermal Hall transport is an interference between impurity skew scattering channels with opposite parity. The thermal Hall conductivity $\sim T^{d+2}$ at low $T$ in $d$ dimensions, and has a window of $T$-independent behavior for $T > T_{\rm imp}$, where $T_{\rm imp}$ is determined by the ratio of scattering potentials with opposite parity. We also consider the role of non-specular scattering off the sample boundary, and find that it leads to negligible corrections to thermal Hall transport at low $T$.
\end{abstract}

\maketitle

\tableofcontents

\section{Introduction}
\label{sec:intro}

Recent experiments \cite{Hess19,Kasahara18,GG19,Ong19,Behnia20,GG20,Tanaka20,Boulanger20} have focused renewed interest on thermal Hall effect of correlated electron materials. In some compounds, including the cuprates, it has been argued \cite{Behnia20,kivelson20,GG20,Boulanger20} that the dominant contributions to the thermal Hall conductivity arises from phonons.
Two important theoretical questions arise when computing the thermal Hall conductivity of phonons. First, what is the origin of the `chirality' of the phonons {\it i.e.\/} the breaking of the time-reversal and mirror symmetries, but not their product? Second, given chiral phonons, what is their thermal Hall conductivity? This paper will address the second question.

To set the stage, let us briefly discuss the first question. As phonons are electrically neutral, any chirality in the phonon Hamiltonian must ultimately arise from their coupling to the electrons. For the cuprates, the enhanced thermal Hall effect is limited to the underdoped regime, implying that the electronic chirality is connected to the novel strong correlation physics of the pseudogap phase \cite{GG20,Boulanger20}. There have been theoretical proposals for the origin of electronic chirality in the pseudogap \cite{SamajdarThermal19,LeeP19,SamajdarEnhanced19,LeeDH19,TaoLi19,GuoGauge20,Varma20},
and de la Torre {\it et al.} \cite{Hsieh20} have noted a connection to recent optical second harmonic generation experiment. Given chiral electrons, then the electron-phonon coupling is known to induce non-dissipative phonon Hall viscosity terms in the effective action for the phonons \cite{Barkeshli12,Ryu15,Landsteiner15,Asgari19,kivelson20,Perkins20,Lucas20}. For the square lattice case relevant to the cuprates, the phonon Hall viscosity induced by a model of chiral spinons \cite{SamajdarEnhanced19} is described in a separate paper \cite{ZhangTeng21}.

Now we can turn to the second question above, which will be addressed by us in this paper: given a phonon system with a non-zero Hall viscosity, what is its thermal Hall conductivity?
This question has not received significant attention in the literature. By analogy with computations of the anomalous Hall effect of electrons \cite{NagaosaRMP,Sinova07}, and as argued by Chen {\it et al.} \cite{kivelson20}, we can separate the contributions to the phonon thermal Hall conductivity into instrinsic and extrinsic terms. The intrinsic constribution is present in a perfect infinite crystal without impurities, and is a consequence of the Berry curvature in the phonon band structure arising from the Hall viscosity term in the phonon Hamiltonian: an explicit formula relating the intrinsic thermal Hall conductivity to the phonon Hall viscosity was obtained in Refs.~\cite{Shi12,kivelson20}. However, Chen {\it et al.} \cite{kivelson20} also argued that this instrinsic thermal Hall effect is too small to explain observations \cite{GG19,Behnia20,GG20,Boulanger20}, and a much larger contribution can arise from extrinsic terms which are inversely proportional to the density of impurities. Chen {\it et al.\/} \cite{kivelson20} made estimates of this extrinsic contribution to the phonon thermal Hall conductivity (which we review below), and we will present here the results of a more complete computation for scattering off point-like impurities. More precisely, the impurity size has to be smaller than the wavelength of the phonons, and this is a mild restriction for acoustic phonons at low temperatures.
Our results do depend inversely on the density of impurities as pointed out by Chen {\it et al.} \cite{kivelson20}, but the proportionality factors are at variance with their estimates for the cases we consider.

Following Chen {\it et al.} \cite{kivelson20}, we will study
the phonon thermal Hall effect from skew scattering on lattice disorder. The skewness is induced by the phonon Hall viscosity. The theory we study
has the Lagrangian density
\begin{equation}\label{eq:Ltot}
  \mathcal{L}=\mathcal{L}_{ph}+\mathcal{L}_H+\mathcal{L}_{dis}.
\end{equation}
Here $\mathcal{L}_{ph}$ is the elastic theory of phonons in a tetragonal lattice; $\mathcal{L}_H$ denotes the phonon Hall viscosity term; $\mathcal{L}_{dis}$ describes lattice disorder from point-like impurities. The explicit forms of these terms will be presented in Section~\ref{sec:model}. Note that all terms in $\mathcal{L}$ are quadratic in the phonon displacement co-ordinate $u^i$, and so the problem is ultimately one of harmonic oscillators in the presence of disorder. Nevertheless, computation of the thermal Hall effect has numerous subtleties, as we shall describe.

Chen {\it et al.} \cite{kivelson20} assumed that the non-skew scattering comes from grain boundary scattering with a mean-free time $\tau$ independent of phonon energy, and the skew-scattering arises from a coupling to Berry curvature $\Omega(k)$, to yield an impurity scattering amplitude of the from
\begin{equation}\label{eq:gammaAOmega}
  \gamma^A(k,k')=\frac{A}{\tau}\Omega(k)\cdot(\vec{k}\times \vec{k}'),
\end{equation}
where $A$ is a prefactor independent of phonon energy. Plugging these into the phonon Boltzmann equation, they found in 3+1D that the low temperature longitudinal and the Hall thermal conductivities are $\kappa_L\propto \ell T^3$ and $\kappa_H\propto \ell T^4$, where $T$ is temperature, $\ell=w\tau$ is the mean-free path with $w$ an acoustic phonon velocity. In a system with dilute disorder, $\ell \propto 1/n_i$ where $n_i$ is disorder density. This $1/n_i$ enhancement is proposed to explain the large thermal Hall observed in experiments.

Section~\ref{sec:model} will introduce the model of phonons with a non-zero Hall viscosity, and their coupling to impurities.
We will compute the non-skew and skew scattering rates from \eqref{eq:Ltot}
in Section~\ref{sec:scatter}, and then insert them into the Boltzmann equation to compute the thermal Hall effect in Section~\ref{sec:thermalHall}.
We confirm the $1/n_i$ enhancement of Chen {\it et al.} \cite{kivelson20}, but not their temperature dependence. We find that the skew scattering rate $\gamma^A$ can be decomposed into even-parity and odd-parity channels. While the even-parity channel does scale as $k^2$, it does not contribute to the thermal Hall effect because of parity considerations. The thermal Hall effect is proportional to skew scattering from the odd-parity channel, which according to our power-counting will scale as $k^{3+d}$ at low momenta, where $d$ is the spatial dimension. Therefore \eqref{eq:gammaAOmega} overestimates the thermal Hall effect at very low temperatures when applied to point-like impurities.

Our main qualitative estimates for the thermal Hall effect from skew scattering of phonons appear in Section~\ref{sec:general}, and complete quantitative computations in two- and three-dimensional crystals are in Section~\ref{sec:2D}
and~\ref{sec:3D} respectively. Apart from the lowest temperature regimes just discussed, we find a crossover above a temperature $T_{\rm imp} \sim w \sqrt{\left|{a}/{b}\right|}$ (see \eqref{eq:Timp}), above which the thermal Hall conductivity is temperature independent. Here $a$ and $b$ are the couplings associated with the coupling of phonons to impurities defined in \eqref{eq:Vdis}. Note that we are assuming $T_{\rm imp} \ll T_D$, where $T_D$ is the Debye temperature, and the temperature independent $\kappa_H$ is for the
$T_{\rm imp} \ll T \ll T_D$.

Section~\ref{sec:boundary} considers the role of sample boundaries in thermal Hall transport. At low $T$, the phonon mean free path can become comparable to the sample size, and then boundary scattering can play an important role in thermal transport. Our analysis shows that the influence of the sample boundary is significantly weaker for Hall transport than for longitudinal transport.

\section{The model}
\label{sec:model}

  In this section we will describe details of the model \eqref{eq:Ltot}. We shall describe the model in three spatial dimensions {\it i.e.\/} (3+1)D, but in later sections we will also consider it in (2+1)D, by dropping the $z$-direction.

\subsection{The elastic theory of phonons}

    The dynamical variables are the displacement fields $u^{i}$ with three components $i=x,y,z$ or $i=1,2,3$. The elastic phonon Lagrangian takes the form
\begin{equation}\label{eq:Lph}
  \mathcal{L}_{ph}=T-U,
\end{equation} where $T$ is the kinetic energy and $U$ is the elastic potential. The kinetic energy takes the conventional form
\begin{equation}\label{}
  T=\frac{\rho}{2}(\partial_t u^i)^2,
\end{equation}with $\rho$ being the mass density.

  To describe the elastic potential $U$, we need to use the strain tensor and strain components \cite{AshcroftNeilW1976Ssp}:
\begin{equation}
  u_{ij}=\frac{1}{2}(\partial_i u^j+\partial_j u^i),
\end{equation}
\begin{equation}
  e_{ij}=\left\{
           \begin{array}{ll}
             u_{ij}, & i=j; \\
             2u_{ij}, & i\neq j.
           \end{array}
         \right.
\end{equation}
  We also introduce short hands for the double index $ij$:
$$
xx\to 1,~yy\to 2,~zz\to 3,~yz\to 4,~zx\to 5,~xy \to 6.
$$

The elastic potential is
\begin{equation}\label{}
  U=\frac{1}{2}\sum_{\alpha\beta}^6\int \rd^3 x e_{\alpha}(x)C_{\alpha\beta}e_\beta(x).
\end{equation} Here the coefficients $C_{\alpha\beta}$ are elastic constants. With applications to cuprates in mind, we will consider tetragonal crystals, with six nonzero elastic constants $C_{11}=C_{22}$, $C_{12}$, $C_{13}$, $C_{33}$, $C_{44}=C_{55}$, $C_{66}$.

We can rewrite the elastic potential in terms of displacement fileds $u^i$ in fourier space as
\begin{equation}\label{}
  S_U=-\frac{1}{2}\int \frac{\rd^4 k}{(2\pi)^4}u^i(-k)K^{ij}(k)u^j(k),
\end{equation}where
\begin{equation}\label{eq:Kmat}
  K^{ij}(k)=\left(
\begin{array}{ccc}
 C_{11} k_x^2+C_{66} k_y^2+C_{44} k_z^2 & \left(C_{12}+C_{66}\right) k_x k_y & \left(C_{13}+C_{44}\right) k_x k_z \\
 \left(C_{12}+C_{66}\right) k_x k_y & C_{66} k_x^2+C_{11} k_y^2+C_{44} k_z^2 & \left(C_{13}+C_{44}\right) k_y k_z \\
 \left(C_{13}+C_{44}\right) k_x k_z & \left(C_{13}+C_{44}\right) k_y k_z & C_{44} \left(k_x^2+k_y^2\right)+C_{33} k_z^2 \\
\end{array}
\right).
\end{equation}

\subsection{Phonon Hall viscosity}

In our model, the phonon Hall viscosity \cite{Barkeshli12} serves as the source of time-reversal breaking and skew scattering. It is the lowest order time-reversal breaking term for phonons in the effective field theory sense. As discussed in Section~\ref{sec:intro}, it can be obtained by coupling lattice distortions to an electronic chiral spin liquid, and then integrating out the electrons \cite{ZhangTeng21}.

The Hall viscosity term can be written as
\begin{equation}\label{eq:LH1}
  \mathcal{L}_H=2\left[\eta^H(u_{xx}-u_{yy})\partial_t u_{xy}+\eta^{M}(u_{xx}+u_{yy})\partial_t m_{xy}\right],
\end{equation}where $m_{xy}=(1/2)(\partial_x u^y-\partial_y u^x)$.
Note that we only include Hall viscosity terms in the $x$-$y$ plane, assuming any applied magnetic field is oriented in the $z$ direction. In such a model, the thermal Hall co-efficients $\kappa_{xz}$ and $\kappa_{yz}$ will vanish because of mirror symmetry across the $x$-$y$ plane. As we will be working to linear order in the Hall viscosity, the thermal Hall conductivity for other field orientations can be determined simply the adding the contributions for the fields along the co-ordinate axes.
Using integration by parts, \eqref{eq:LH1} can be rewritten as
\begin{equation}\label{eq:SH}
  S_H=\int \rd^4 x \left(\frac{-\eta}{2}\right)\left((\partial_x^2+\partial_y^2)u^x \partial_t u^y-(\partial_x^2+\partial_y^2)u^y \partial_t u^x\right),
\end{equation}where $\eta=\eta^H+\eta^M$ is the Hall viscosity. Converting to fourier space, this is
\begin{equation}\label{}
  S_H=\int\frac{\rd ^4 k}{(2\pi)^4}(\eta^H+\eta^M)\frac{-i\omega}{2}(k_x^2+k_y^2)(u^x(-k)u^y(k)-u^y(-k)u^x(k)).
\end{equation} In the rest of the paper, we will treat the phonon Hall viscosity perturbatively, to first order in $\eta$.

\subsection{Quantizing free phonons}

  Now we quantize the phonon action $S=\int \rd^4 x ( \mathcal{L}_{ph}+\mathcal{L}_H )$ in absence of disorder. The goal is to identify the creation and annihilation operators. To carry out canonical quantization, first we find the generalized momentum
\begin{eqnarray}
  \pi^x&=&\frac{\delta S}{\delta \partial_t u^x}=\rho \partial_t u^x+\frac{\eta}{2}(\partial_x^2+\partial_y^2)u^y,\\
 \pi^y&=&\frac{\delta S}{\delta \partial_t u^y}=\rho \partial_t u^y-\frac{\eta}{2}(\partial_x^2+\partial_y^2)u^x,\\
  \pi^z &=& \frac{\delta S}{\delta \partial_t u^z}=\rho \partial_t u^z.
\end{eqnarray} The Hamiltonian is
\begin{equation}\label{}
  \mathcal{H}=\frac{1}{2\rho}\left[(\pi^x-\frac{\eta}{2}\nabla^2u^y)^2+(\pi^y+\frac{\eta}{2}\nabla^2 u^x)^2+(\pi^z)^2\right]+\frac{1}{2}u^iK^{ij}u^j,
\end{equation} where $\nabla^2=\partial_x^2+\partial_y^2$.

To diagonalize the Hamiltonian, we follow \cite{kivelson20}, by first grouping the canonical variables
\begin{equation}\label{}
  \zeta^I=(u^i,\pi_i),\quad I=1\dots6,
\end{equation}which admits a symplectic structure
\begin{equation}\label{}
  [\zeta^I,\zeta^J]=i J^{IJ}.
\end{equation} We are using the same notation as \cite{kivelson20}, where lower $i=1,2,3$ denotes momentum and upper $i=1,2,3$ denotes displacement.
The Hamiltonian then has a matrix representation (by hermiticity we have $\zeta^{I}(k)=\zeta^{I}(-k)^\dagger$)
\begin{equation}\label{eq:Hk3}
  H=\frac{1}{2}\int\frac{\rd^3 k}{(2\pi)^3}\zeta^I(k)^\dagger H_{IJ}(k)\zeta^{J}(k),
\end{equation}where we can organize the Hamiltonian $H=H_0+H_1+H_2$ in powers of $\eta$ and
\begin{equation}\label{}
 H_0=\left(
\begin{array}{cccccc}
 C_{11} k_x^2+C_{66} k_y^2+C_{44} k_z^2 & \left(C_{12}+C_{66}\right) k_x k_y & \left(C_{13}+C_{44}\right) k_x k_z & 0 & 0 & 0 \\
 \left(C_{12}+C_{66}\right) k_x k_y & C_{66} k_x^2+C_{11} k_y^2+C_{44} k_z^2 & \left(C_{13}+C_{44}\right) k_y k_z & 0 & 0 & 0 \\
 \left(C_{13}+C_{44}\right) k_x k_z & \left(C_{13}+C_{44}\right) k_y k_z & C_{44} \left(k_x^2+k_y^2\right)+C_{33} k_z^2 & 0 & 0 & 0 \\
 0 & 0 & 0 & \frac{1}{\rho } & 0 & 0 \\
 0 & 0 & 0 & 0 & \frac{1}{\rho } & 0 \\
 0 & 0 & 0 & 0 & 0 & \frac{1}{\rho } \\
\end{array}
\right),
\end{equation}
\begin{equation}\label{}
  H_1=\left(
\begin{array}{cccccc}
 0 & 0 & 0 & 0 & -\frac{\eta  \left(k_x^2+k_y^2\right)}{2 \rho } & 0 \\
 0 & 0 & 0 & \frac{\eta  \left(k_x^2+k_y^2\right)}{2 \rho } & 0 & 0 \\
 0 & 0 & 0 & 0 & 0 & 0 \\
 0 & \frac{\eta  \left(k_x^2+k_y^2\right)}{2 \rho } & 0 & 0 & 0 & 0 \\
 -\frac{\eta  \left(k_x^2+k_y^2\right)}{2 \rho } & 0 & 0 & 0 & 0 & 0 \\
 0 & 0 & 0 & 0 & 0 & 0 \\
\end{array}
\right),
\end{equation}
\begin{equation}\label{}
H_2=\left(
\begin{array}{cccccc}
 \frac{\eta ^2 \left(k_x^2+k_y^2\right){}^2}{4 \rho } & 0 & 0 & 0 & 0 & 0 \\
 0 & \frac{\eta ^2 \left(k_x^2+k_y^2\right){}^2}{4 \rho } & 0 & 0 & 0 & 0 \\
 0 & 0 & 0 & 0 & 0 & 0 \\
 0 & 0 & 0 & 0 & 0 & 0 \\
 0 & 0 & 0 & 0 & 0 & 0 \\
 0 & 0 & 0 & 0 & 0 & 0 \\
\end{array}
\right).
\end{equation}
We can diagonalize the matrix $iJH(k)$
\begin{equation}\label{eq:SymplecticDiag1}
  M(k)iJH(k)M(k)^{-1}=\ce(k),
\end{equation} or
\begin{equation}\label{eq:SymplecticDiag2}
  M(k)^A_{~I}(iJH(k))^{I}_J=\ce^{A}_{~B}(k) M^B_{~J}(k),
\end{equation}where $\ce^A_{~B}={\rm diag}(E^\alpha(k),-E^{\alpha}(-k))$.
It is shown in \cite{kivelson20} that we can normalize $M(k)$ so that
\begin{equation}\label{}
  M^A_{~I}(k)(iJ)^{IJ}M^{B}_{~J}(k)^*=\delta^{AB}\sgn \ce^{A}_{~A}(k),\quad\text{$A$ is not summed}.
\end{equation}

The creation and annihilation operators are obtained as
\begin{equation}\label{eq:chiA}
  \chi^A(k)=\begin{pmatrix}
              a^\alpha(k) \\
              a_\beta(-k)^\dagger
            \end{pmatrix}=M^A_{~I}(k)\zeta^I(k)
\end{equation}
We can write down $M^A_I$ in terms of block matrices
\begin{equation}\label{}
  M^A_{~I}(k)=\begin{pmatrix}
              M^\alpha_{~I}(k) \\
              \delta_{\beta\beta'}M^{\beta'}_{~I}(-k)^*
            \end{pmatrix}=
\begin{pmatrix}
  (M_u)^\alpha_{~i}(k) & \delta^{\alpha\alpha'}(M_\pi)_{\alpha'}^{~j}(k) \\
  \delta_{\beta\beta'}(M_u)^{\beta'}_{~i}(-k)^* & (M_\pi)_{\beta}^{~j}(-k)^*
\end{pmatrix}.
\end{equation} When $\eta=0$, $M_u$ is a real and $M_\pi$ is pure imaginary (entry-wise). Since the Hamiltonian is even in $k$, we have $M(k)=M(-k)$.
At zeroth order in $\eta$, we can write down $M_0$ as
\begin{equation}\label{eq:M0k}
  M_0(k)=\frac{1}{\sqrt{2}}
\begin{pmatrix}
  I_3   & i I_3 \\
  I_3 & -i I_3
\end{pmatrix}
\begin{pmatrix}
  \begin{pmatrix}
    \sqrt{\rho E^1(k)} (e_k^1)^T \\
    \sqrt{\rho E^2(k)} (e_k^2)^T \\
    \sqrt{\rho E^3(k)} (e_k^3)^T
  \end{pmatrix} & 0 \\
  0 & \begin{pmatrix}
        \frac{1}{\sqrt{\rho E^1(k)}} (e_k^1)^T \\
        \frac{1}{\sqrt{\rho E^2(k)}} (e_k^2)^T \\
        \frac{1}{\sqrt{\rho E^3(k)}} (e_k^3)^T
      \end{pmatrix}
\end{pmatrix},
\end{equation} where $I_3$ denotes 3 by 3 identity matrix, and $e_k^\alpha$ is a column vector which describes the polarization of the $\alpha$-th phonon band. It can be obtained by solving the eigenvalue problem
\begin{equation}\label{}
  K^{ij}(k)(e_k^\alpha)_j=\rho (E^\alpha(k))^2 (e_k^\alpha)_i,
\end{equation} where $K^{ij}$ is defined by \eqref{eq:Kmat}. A fact that will be useful later is that for generic tetragonal crystal, the three phonon bands are non-degenerate except on the $k_z$ axes.

Now we develop a perturbation theory for $M(k)$ to first order in $\eta$. Let's write
\begin{equation}\label{eq:M1def}
  \begin{split}
     H & = H_0+H_1, \\
     M  &=(1+M_1)M_0,\\
     \ce &=\ce_0+\ce_1,
  \end{split}
\end{equation} such that $M_0$ diagonalize $iJH_0$ to diagonal matrix $\ce_0$. We assume that $\ce_1$ is diagonal.
To linear order, we have
\begin{equation}\label{eq:ce1}
  \ce_1=[M_1,\ce_0]+M_0(iJH_1)M_0^{-1}.
\end{equation} Taking diagonal components of \eqref{eq:ce1}, we have
\begin{equation}\label{}
  (\ce_1)^A_{~A}=(M_0(iJH_1)M_0^{-1})^A_{~A},
\end{equation} and the off-diagonal component yields
\begin{equation}\label{eq:M1AB}
  (M_1)^A_{~B}=\frac{1}{(\ce_0)^A_{~A}-(\ce_0)^B_{~B}}(M_0(iJH_1)M_0^{-1})^A_{~B},
\end{equation} and as in usual perturbation theory we assume $M_1$ has no diagonal entries. The perturbation theory is well defined even at the seemingly degenerate $k_z$-axes. This is because the denominator of
\eqref{eq:M1AB} vanishes linearly as $\sqrt{k_x^2+k_y^2}$, but the numerator vanishes quadratically as $k_x^2+k_y^2$.
For later computations, we will also need the phonon velocity, which is given by
\begin{equation}\label{}
  v^\alpha_i=\frac{\partial E^\alpha(k)}{\partial k_i}=(M(iJ\frac{\partial H(k)}{\partial k_i})M^{-1})^\alpha_{~\alpha}.
\end{equation}
\subsection{Disorder term}

  At last we discuss the disorder term due to impurities, which has the form
\begin{equation}\label{eq:Ldis}
  \mathcal{L}_{dis}=-n_{imp}(x)V_{dis}[u].
\end{equation} Here $n_{imp}(x)$ is the impurity density and $V_{dis}[u]$ describes the phonon-impurity coupling. By translation symmetry, $V_{dis}[u]$ only depends on derivatives of $u^i$. For this paper, we will be focusing on
\begin{equation}\label{eq:Vdis}
  V_{dis}[u]=a (\partial_i u^j)^2+b(\partial^2 u^j)^2+\dots.
\end{equation} The result of this paper is that we need both $a,b\neq 0$ to get a thermal Hall effect from skew scattering. One can also write down other terms like $(\partial_t u^i)^2$, $(\partial_i u^i)^2$ etc., but we found there is no essential difference to the physics.

  For the impurity density, we assume it describes a set of independent point impurities, with
\begin{equation}\label{eq:rho_imp}
  n_{imp}(x)=\sum_a \delta(x-x_a),
\end{equation}where $x_a$'s are independent random positions. The fourier transform is
\begin{equation}\label{}
  n_{imp}(q)=\sum_a e^{-iq\cdot x_a}.
\end{equation}
Performing disorder average of $n_{imp}(q)$ up to cubic order, we obtain
\begin{equation}\label{}
  \overline{n_{imp}(q)}=n_i(2\pi)^3\delta(q);
\end{equation}
\begin{equation}\label{}
  \overline{n_{imp}(q_1)n_{imp}(q_2)}=n_i^2 (2\pi)^6 \delta(q_1)\delta(q_2)+n_i(2\pi)^3 \delta(q_1+q_2);
\end{equation}
\begin{equation}\label{eq:rhoi3}
\begin{split}
  &\overline{n_{imp}(q_1)n_{imp}(q_2)n_{imp}(q_3)}=n_i^3 (2\pi)^9 \delta(q_1)\delta(q_2)\delta(q_3)\\
&+n_i^2(2\pi)^6\left(\delta(q_1+q_2)\delta(q_3)+\delta(q_1+q_3)\delta(q_2)+\delta(q_2+q_3)\delta(q_1)\right)+n_i (2\pi)^3 \delta(q_1+q_2+q_3).
\end{split}
\end{equation} Here $n_i=N/V_s$ is the disorder density, $V_s$ being the spatial volume. We will assume $n_i$ is small enough so we can ignore correction to mass density $\rho$ from $\overline{n_{imp}}$.

\section{Scattering Rate}\label{sec:scatter}

  In this section we describe the skew scattering rate of phonon. We shall first work out the effective scattering potential on a single phonon, and then write down the skew-scattering rate using Born's approximation \cite{Sinova07}.

\subsection{Single phonon scattering potential}\label{sec:Vsp}

We write the disorder term \eqref{eq:Ldis} in Hamiltonian form
\begin{equation}\label{eq:Qdef}
\begin{split}
  H_{dis}&=\int \rd^3 x n_{imp}(x)V_{dis}[u]\\
        &= \int\frac{\rd^3 p\rd^3 q}{(2\pi)^6}n_{imp}(q-p)\zeta^J(q)^\dagger Q_{JI}(q,p)\zeta^I(p).
\end{split}
\end{equation}For our choice of $V_{dis}$ in \eqref{eq:Vdis}, the $Q$-matrix takes the form
\begin{equation}\label{eq:Q3}
  Q_{JI}(q,p)=\begin{pmatrix}
                a\vec{p}\cdot \vec{q}+b \vec{p}^2\vec{q}^2 & 0 \\
                0 & 0
              \end{pmatrix}\otimes I_3,
\end{equation} where $I_3$ is 3 by 3 identity matrix.

  We can transform to the basis of creation and annihilation operators, yielding
\begin{equation}\label{}
  H_{dis}=\int\frac{\rd^3 p\rd^3 q}{(2\pi)^6}n_{imp}(q-p)\chi^B(q)^\dagger P_{BA}(q,p)\chi^A(p),
\end{equation}where
\begin{equation}\label{eq:Pmat}
  P_{BA}(q,p)=\left[(M^{-1}(q)^\dagger Q(q,p) M^{-1}(p))\right]_{BA}.
\end{equation}
  Picking out contributions containing only $a^\dagger a$, we have
\begin{equation}\label{}
  H_{dis}\supset\int\frac{\rd^3 p\rd^3 q}{(2\pi)^6}n_{imp}(q-p)a_{\alpha'}^\dagger(q)a^\alpha(p)\left[P^{\alpha'}_{~\alpha}(q,p)+P_\alpha^{~\alpha'}(p,q)\right].
\end{equation}Here $P^{\alpha'}_{~\alpha}$ is the top-left block and $P_\alpha^{~\alpha'}$ is the bottom-right block in the order of basis in \eqref{eq:chiA}.
  Therefore the matrix element of the single-particle scattering potential $V_{sp}$ is
\begin{equation}\label{eq:Fdef}
  \braket{q,\beta|V_{sp}|p,\alpha}=\frac{1}{V_s}n_{imp}(q-p)\left[P^\beta_{~\alpha}(q,p)+P_\alpha^{~\beta}(p,q)\right]\equiv \frac{1}{V_s}n_{imp}(q-p)F_{\beta\alpha}(q,p).
\end{equation} Here $V_s$ on the RHS is system volume, and $(p,\alpha)$ labels the phonon momentum and band index.  When $\eta=0$, the above matrix element is real. By construction the $F$ matrix is hermitian: $F_{\alpha\beta}(p,q)=F_{\beta\alpha}(q,p)^*$ .

\subsection{Scattering rate}
 In this section we compute both the non-skew and skew scattering rates. We shall use $l$ to label the single phonon states, $E_l$ to denote the single particle energy and $k_l$ to denote the momentum. The scattering rate is given by Fermi's golden rule
\begin{equation}\label{eq:FGrule}
  \gamma_{ll'}=2\pi\overline{|T_{ll'}|^2}\delta(E_l-E_{l'}),
\end{equation} where the $T$-matrix is block diagonal in energy and is given by Lippmann-Schwinger equation
\begin{equation}\label{eq:LS}
  T(E)=V_{sp}+V_{sp}\frac{1}{E-H_0+i\epsilon}T(E),
\end{equation}and $\overline{|T_{ll'}|^2}$ means disorder average.

The leading order term is symmetric under $l\leftrightarrow l'$ and contributes to non-skew scattering rate: $(k_l\neq k_{l'})$
\begin{equation}\label{eq:gammaS}
  \gamma^S_{ll'}=\frac{2\pi n_i}{V_s}|F_{ll'}(k_l,k_{l'})|^2\delta(E_l-E_{l'}).
\end{equation} Here the superscript $S$ means symmetric. There is also a forward-scattering term of order $n_i^2$ that we have dropped, because it is subdominant in $n_i$ and it doesn't contribute to transport.
According to \cite{Sinova07}, the lowest order contribution to skew scattering comes from cubic order in $V_{sp}$:
\begin{equation}\label{}
  \gamma_{ll'}=-(2\pi)^2\sum_{l''}\Im\overline{V_{sp,ll'}V_{sp,l'l''}V_{sp,l''l}}  \delta(E_l-E_{l'})\delta(E_l-E_{l''})\,.
\end{equation}
Taking disorder average using \eqref{eq:rhoi3}, there will be three contributing terms. The first term of three delta functions yields
\begin{eqnarray}\label{}
  \gamma_{ll'} &\supset& -\frac{(2\pi)^2n_i^3}{V_s}(2\pi)^3\delta(k_l-k_{l'})\sum_{l''}\Im\left[F_{ll''}(k_l,k_l)F_{l''l'}(k_l,k_l)F_{l'l}(k_l,k_l)\right] \nonumber \\
  &~&~~~~~~~~~\times \delta(E_l(k_l)-E_{l'}(k_{l'}))\delta(E_l(k_l)-E_{l''}(k_l)).
\end{eqnarray}Here the momentum integration has been performed and the sum runs over band indices. Since for generic $k$, the three bands are non-degenerate, the energy delta functions will set $l=l'=l''$, and $F_{ll}(k,k)$ is real by hermiticity and therefore this term vanishes.
The second term which contains two $\delta$-functions is
\begin{equation}\label{}
\begin{split}
   \gamma_{ll'}&\supset-\frac{(2\pi)^2n_i^2}{V_s}\delta(E_l(k_l)-E_{l'}(k_{l'}))\sum_{l''} \\
     & \Im \Big[(2\pi)^3\delta(k_l-k_{l'})\int \frac{\rd^3 k_{l''}}{(2\pi)^3}F_{ll''}(k_l,k_{l''})F_{l''l'}(k_{l''},k_l)F_{l'l}(k_l,k_l)\delta(E_l(k_l)-E_{l''}(k_{l''}))\\
& +F_{ll''}(k_l,k_{l'})F_{l''l'}(k_{l'},k_l')F_{l'l}(k_{l'},k_l)\delta(E_l(k_l)-E_{l''}(k_{l'}))\\
&+ F_{ll''}(k_l,k_l)F_{l''l'}(k_l,k_l')F_{l'l}(k_{l'},k_l)\delta(E_l(k_l)-E_{l''}(k_l))\Big].
\end{split}
\end{equation}In the second line, the energy delta function imposes $l=l'$ and the integral is explicitly real. In the third line, the energy delta function imposes $l'=l''$, and then it's real. In the forth line, the delta function imposes $l=l''$, and the term is real.
Therefore only the linear in $n_i$ term contributes to skew-scattering
\begin{eqnarray}\label{eq:gammaA}
  \gamma^{A}_{ll'} &=& -\frac{(2\pi)^2 n_i}{V_s}\delta(E_l(k_l)-E_{l'}(k_{l'}))\sum_{l''}\int\frac{\rd^3 k_{l''}}{(2\pi)^3} \nonumber \\ &~&~~~~~~~~~~~\Im\left[F_{ll''}(k_l,k_{l''})F_{l''l'}(k_{l''},k_{l'})F_{l'l}(k_{l'},k_l)\right]\delta(E_l(k_l)-E_{l''}(k_{l''})).
\end{eqnarray} Here the superscript $A$ means antisymmetric.

  Finally, we remark that there is a cubic in $V_{sp}$ correction to the non-skew scattering rate $\gamma^S$, but it can be safely ignored to linear order in $\eta$.

\section{The thermal Hall effect}\label{sec:thermalHall}

In this section we shall compute the thermal Hall effect using the Boltzmann equation approach.
\subsection{Botlzmann equation}
   Under a temperature gradient $\nabla T$, the Boltzmann equation around equilibrium takes the form
\begin{equation}\label{eq:Boltzmann}
   -\frac{\partial n_B}{\partial E_l}\frac{E_l \vec{v}_l}{T}\cdot\nabla T=-I[f_l],
\end{equation}where the collision integral $I[f_l]$ is
\begin{equation}\label{eq:If}
  I[f_l]=\sum_{l'}\left(\gamma_{l'l}f_{l}-\gamma_{ll'}f_{l'}\right).
\end{equation} As in previous sections, $l$ labels single particle states of phonons. According to \cite{KL1957}, the collision term is linear in $f_l$ rather than $f_l(1+f_{l'})$. The absence of Bose enhancement is related to the fact that scattering in impurity potential is ultimately a one-body problem, so many-body statistics is not relevant.

Since energy is conserved during scattering, we can consider solving \eqref{eq:Boltzmann} with fixed energy $E$, i.e. consider the equation
\begin{equation}\label{eq:gli}
  v_{li} \delta(E_l-E)=I[g_{li}(E)],
\end{equation} here the additional index $i=x,y,z$ denote three components of the velocity. The relation between $f_l$ and $g_{li}$ is
\begin{equation}\label{}
  f_l=\int_0^{\infty} \rd E \frac{\partial n_B}{\partial E} \frac{E}{T} g_{li}(E) \partial_i T.
\end{equation} Using the definition of heat current
\begin{equation}\label{}
  \vec{j_Q}=\frac{1}{V_s}\sum_{l}\vec{v}_l E_l f_l,
\end{equation}we can write down the thermal conductivities in a spectral representation as
\begin{equation}\label{eq:kappaK}
  \kappa_{ij}=\int_0^{\infty} \rd E \left(-\frac{\partial n_B}{\partial E}\right) \frac{E^2}{T} \cK_{ij}(E),
\end{equation}where
\begin{equation}\label{eq:calK}
  \cK_{ij}(E)=\frac{1}{V_s}\sum_{E_l=E} v_{li}g_{lj}(E)
\end{equation} is referred as spectral thermal conductivity. Because of the $\delta$-function in \eqref{eq:gli}, the sum here only includes states with energy $E_l=E$.

We can also write down $\cK$ using functional notation, as
\begin{equation}\label{}
  \cK_{ij}(E)=\left\langle v_i(E), I^{-1} v_j(E)\right\rangle.
\end{equation}  Here we view the velocity $v_i(E)$ as a function on the states with energy $E_l=E$ and the collision integral $I$ as a linear functional acting on this space. The action of $I$ is given in \eqref{eq:If}, and the inner product is defined as
\begin{equation}\label{}
  \braket{F,G}=\frac{1}{V_s} \sum_{l,E_l=E} F_l G_l.
\end{equation}
We should point out that the collision operator $I$ has a zeromode $g^{(0)}_l=1$\footnote{\label{fn:1}To show this, we shall verify $\sum_{l'}\gamma_{ll'}=\sum_{l'}\gamma_{l'l}$. By optical theorem, both sides are equal to the imaginary part of forward $l\to l$ scattering amplitude. As a corollary the sum over antisymmetric part of $\gamma_{ll'}$ vanishes identically.}, and therefore $I^{-1}v_j$ is ambiguous by $g^{(0)}$, but this ambiguity can be ignored because it physically corresponds to the equilibrium solution and doesn't contribute to transport.

Using this functional notation, we can conveniently perform a perburbative expansion in $\eta$. We can write the collision operator as the sum of non-skew scattering and skew-scattering contributions
\begin{equation}\label{}
  I=I_S+I_A,
\end{equation} where $I_S$ only involves non-skew scattering $\gamma^{S}_{ll'}$ and $I_A$ only involves skew scattering $\gamma^{A}_{ll'}$. Here $I_A$ is proportional $\eta$, and to first order in $\eta$ we have
\begin{equation}\label{eq:Kijfunc}
  \cK_{ij}(E)=-\braket{v_i(E),I_S^{-1}I_A I_S^{-1}v_j(E)}.
\end{equation} We will use this to carry out symmetry analysis in the next part.

\subsection{General analysis of the thermal Hall effect}
\label{sec:general}

  We now argue based on parity symmetry that the thermal Hall effect originating from skew-scattering is quite small in the presence of a single impurity scattering channel. The impurity potential (\ref{eq:Vdis}) contains two scattering channels, with co-efficients $a$ and $b$, and both are needed to obtain a skew-scattering Hall effect to linear order in the Hall viscosity.

  The precise statement is the following.
The skew scattering thermal Hall effect is of order $\mathcal{O}(\eta^3)$ or higher under the following assumptions:
\begin{enumerate}
  \item The phonon bands are non-degenerate for generic $k$. As a consequence the individual phonon dispersions will be even under parity. For example, in a tetragonal crystal the phonon bands are non-degenerate and are even under parity (i.e. parity is not spontaneously broken).  A counterexample is the isotropic crystal where the degeneracy between two transverse modes is lifted by $\eta$ and the resulting circularly polarized bands break parity and time-reversal \footnote{On the Hamiltonian level the parity is still good because it exchanges the two circularly polarized bands.}.
  \item The disorder potential only contain channels of the same parity. Using \eqref{eq:Vdis} as an example. The first term $a(\partial_i u^j)^2$ has odd parity and the second term $b(\partial^2 u^j)^2$ has even parity. In momentum space the first term has the form $a\vec{p}\cdot\vec{q}$, and it flips sign when we fix one of $\vec{p},\vec{q}$ and flip the other. In contrast, the second term in momentum space is of the form $b \vec{p}^2\vec{q}^2$ and it doesn't change sign under single momentum flip. Our statement is therefore $\kappa_{xy}=\mathcal{O}(\eta^3)$ if $ab=0$.
\end{enumerate}

  The proof is the following.

  First, the $F$-matrix defined in \eqref{eq:Fdef} also has channels of the same parity. To go from disorder potential to $F$ matrix, we should multiply some factors related to phonon polarization (see from \eqref{eq:Vdis} to \eqref{eq:Fdef}). Under assumption 1, the phonon polarizations can have the same parity as discussed in the next paragraph.  Therefore the $F$-matrix also has a single parity channel.

Notice that including the Hall viscosity term, the Hamiltonian is even under parity $H(k)=H(-k)$ and non-degenerate, so each polarization can have definite parity. In 2D, we can choose all polarization vectors $e^{\alpha}_k$ to be smooth in $k$ and have odd parity $e^{\alpha}_{-k}=-e^{\alpha}_k$. We can achieve this by starting from an isotropic 2D crystal with $e^1_k=(k_x/k,k_y/k)^T$, $e^2_k=(-k_y/k,k_x/k)^T$ and then smoothly deform the elastic constants while preserving the phonon band gap and parity symmetry. In 3D, it is impossible to construct the polarizations as smooth functions of $k$ since there is no smooth vector field on a sphere. However it is still possible to define a non-smooth polarization field with even parity. The non-smoothness of the polarizations shouldn't be a problem since the sign of polarization vector is a gauge choice and will be squared away in scattering rates. In this argument, it's important for the phonon dispersion to be non-degenerate, otherwise the degeneracy could be split in a parity-breaking manner, as is the case for an isotropic crystal.

 Following from the $F$-matrix, the scattering rate $\gamma_{ll'}$ determined from Fermi's Golden rule \eqref{eq:FGrule} and Lippmann-Schwinger equation \eqref{eq:LS} will have even parity. Although the single particle potential $V_{sp}$ given in \eqref{eq:Fdef} is not invariant under parity due to the impurity density $n_{imp}$, the symmetry will be restored in the scattering rate after disorder averaging.

  The first order in $\eta$ thermal Hall conductivity is given by \eqref{eq:Kijfunc} as a matrix element of $I_S^{-1}I_A I_{S}^{-1}$. The velocities $v_i(E)$ and $v_j(E)$ are odd under parity. The symmetric collision integral $I_S$ preserves the parity of $v_j(E)$. This can be seen by writing
\begin{equation}\label{}
  I_S[f_l]=\frac{f_l}{\tau_l}-\sum_{l'}\gamma^S_{ll'}f_{l'},
\end{equation} where $1/\tau_l=\sum_{l'}\gamma^S_{l'l}$ and $\tau_l$ is even under parity. Since $\gamma^S_{ll'}$ only contains even parity channels and therefore annihilates $v_j(E)$, we have $I_{S}^{-1}[v_{lj}(E)]=\tau_l v_{lj}(E)$ which is still odd under parity. Similarly, we can consider the action of the antisymmetric collision integral
\begin{equation}\label{}
  I_A[f_l]=f_l(\sum_{l'}\gamma^A_{l'l})-\sum_{l'}\gamma^A_{ll'}f_{l'}.
\end{equation} The sum in the parentheses vanishes identically as a consequence of the optical theorem, see footnote \ref{fn:1}. Therefore $I_A$ only contains even-parity channels, and annihilates $I_S^{-1} [v_{j}(E)]$.

  To obtain a thermal Hall conductivity linear in $\eta$, we should break either of the assumptions listed at beginning of this subsection. Degenerate phonon bands are unlikely in the cuprates, so the only option is to break assumption 2 by introducing two scattering channels of different parity. This is exactly what we have written down in \eqref{eq:Vdis}, based on locality and translation symmetry.

  We can perform a rough power counting analysis for the thermal conductivities. The goal is to determine the temperature powers of the $\kappa_{xx}$ and $\kappa_{xy}$ at low and high temperatures.

  To begin with, we notice that the phonon dispersion is not corrected to first order in $\eta$, because the phonon Hall viscosity is time-reversal odd but the zeroth order phonon bands are time-reversal even and non-degenerate. Therefore all momenta are linear in energy, and we can schematically write the disorder potential as
\begin{equation}\label{}
  V_{dis}\sim (a E^2+b E^4)u^2,
\end{equation}where $E$ is the energy of the scattered phonon and we have dropped other factors.
Using \eqref{eq:M0k}, each displacement field $u^i$ contributes energy dimension $-1/2$, and from \eqref{eq:SH} the Hall viscosity $\eta$ has energy dimension -1, so the $F$-matrix will take the form
\begin{equation}\label{}
  F_{ll'}\sim (aE+b E^3)(1+\eta E).
\end{equation} Using \eqref{eq:gammaS}, the scattering rate scales as
\begin{equation}\label{}
  \gamma^S_{ll'}\sim n_i E^2(a+bE^2)^2.
\end{equation} From \eqref{eq:gammaA}, the skew scattering rate is proportional to cube of $F_{ll'}$, and it should also be proportional to $\eta$. From this we have
\begin{equation}\label{}
  \gamma^A_{ll'}\sim n_i \eta E^{d+3}(a+bE^2)^{3},
\end{equation} where the dependence on spatial dimension $d$ comes from summing over intermediate states on the energy shell. As argued before, only the odd-parity channels of $\gamma^A_{ll'}$ contributes to the thermal Hall effect, this corresponds to the cross terms between $a$ and $b$, so the effective skew-scattering rate is
\begin{equation}\label{}
   (\gamma^A_{ll'})_\text{eff}\sim n_iab \eta E^{d+5}(a+b E^2).
\end{equation} We can insert the scattering rates into the Boltzmann equation, and using the fact that velocities do not scale with energy, we have
\begin{equation}\label{}
  \cK_{xx}(E)\sim \frac{1}{\gamma^S}\sim n_i^{-1}E^{-2}(a+bw^{-2} E^2)^{-2} \rho^2 w^6,
\end{equation}
\begin{equation}\label{}
  \cK_{yx}(E)\sim \frac{(\gamma^A)_\text{eff}}{(\gamma^S)^2}\sim n_{i}^{-1}ab \eta E^{d+1}(a+b w^{-2}E^2)^{-3} w^{-d}.
\end{equation} Here the energy powers arising from summing over energy surface cancelled between numerator and denominator. We have also reinstated the sound velocity $w$ (we assumed velocities of all bands are of the same order) and the mass density $\rho$ by dimensional analysis. From the above two expressions we see the emergence of a disorder-related crossover energy/temperature scale
\begin{equation}\label{eq:Timp}
  T_{\rm imp}\sim w \sqrt{\left|\frac{a}{b}\right|}.
\end{equation}

The thermal conductivities can be obtained from \eqref{eq:kappaK}. For the longitudinal thermal conductivity, we found that there is an IR divergence, and at low temperature we have
\begin{equation}\label{eq:kxxrough}
  \kappa_{xx}(T\to 0)\sim \frac{(a+bw^{-2} T^2)^{-2} \rho^2 w^6}{ T(e^{\Delta/T}-1)}n_i^{-1}\,.
\end{equation} where $\Delta$ is the IR energy cutoff due to a finite sample size. This is in agreement with \cite{Stephen83} where they found a similar IR divergence of $\kappa_{xx}$ near (2+1)D, which is a consequence of including only elastic disorder scattering.
The thermal Hall conductivity at low temperature is
\begin{equation}\label{eq:kyxrough}
  \kappa_{yx}(T\to 0)\sim ab \eta n_{i}^{-1}T^{d+2}(a+b w^{-2} T^2)^{-3}w^{-d}.
\end{equation}

We shall emphasize that the results above are only good for power counting. For example, $(a+b w^{-2}E^2)^2$ means that there will be three terms that are proportional to $a^2$, $abw^{-2}E^2$ and $b^2 w^{-4} E^4$ respectively, but the coefficients are to be determined from solving the Boltzmann equation exactly.

 At high temperature (but still below the Debye temperature), the thermal conductivities all saturate to a constant. We can directly take the $T\to\infty$ limit in \eqref{eq:kappaK}, and we obtain
\begin{equation}
    \kappa_{ij}(T\to\infty)=\int_0^\infty \rd E \cK_{ij}(E).
\end{equation} Therefore we have
\begin{eqnarray}
  \kappa_{xx}(T\to \infty) &\sim&  \frac{\rho^2 w^6}{a^2 n_i }\frac{1}{\Delta}\,, \\
  \kappa_{yx}(T\to \infty) &\sim& \frac{b \eta}{a^2 n_i}w^2 \left|\frac{a}{b}\right|^{\frac{d+2}{2}}\,.
\end{eqnarray}

\section{Thermal Hall effect in a 2D isotropic crystal}
\label{sec:2D}

  As a concrete demonstration of the aforementioned results, we explicitly calculate the thermal conductivities in a 2D isotropic crystal.

\subsection{The Hamiltonian}

  For a 2D isotropic lattice with the phonon Hall viscosity term, the Hamiltonian has a matrix representation as in \eqref{eq:Hk3} where $H=H_0+H_1+H_2$, and
\begin{equation}\label{}
  H_0(k)=\begin{pmatrix}
        \mu_1(k_x^2+k_y^2)+\mu_2 k_x^2 & \mu_2 k_x k_y & 0   & 0 \\
       \mu_2 k_x k_y & \mu_1(k_x^2+k_y^2)+\mu_2 k_y^2 & 0 & 0 \\
        0 & 0 & \frac{1}{\rho} & 0 \\
        0 & 0 & 0 & \frac{1}{\rho}
      \end{pmatrix},
\end{equation}
\begin{equation}\label{}
  H_1(k)=\left(
\begin{array}{ccccc}
 0 & 0 &  0 & -\frac{\eta  \left(k_x^2+k_y^2\right)}{2 \rho }  \\
 0 & 0 &  \frac{\eta  \left(k_x^2+k_y^2\right)}{2 \rho } & 0  \\
 0 & \frac{\eta  \left(k_x^2+k_y^2\right)}{2 \rho }  & 0 & 0 \\
 -\frac{\eta  \left(k_x^2+k_y^2\right)}{2 \rho } & 0  & 0 & 0
\end{array}
\right),
\end{equation}
\begin{equation}\label{}
  H_2(k)=\begin{pmatrix}
    \frac{\eta ^2 \left(k_x^2+k_y^2\right){}^2}{4 \rho } & 0 & 0 & 0 \\
    0 & \frac{\eta ^2 \left(k_x^2+k_y^2\right){}^2}{4 \rho } & 0 & 0 \\
    0 & 0 & 0 & 0 \\
    0 & 0 & 0 & 0
  \end{pmatrix}.
\end{equation}

To first order in $\eta$, the dispersion is given by
\begin{equation}\label{}
  E^\alpha(k)=kw_\alpha,\quad \alpha=1,2,
\end{equation}and
\begin{equation}\label{eq:w}
  w_1=\sqrt{\frac{\mu_1+\mu_2}{\rho}},\qquad w_2=\sqrt{\frac{\mu_1}{\rho}}.
\end{equation} The polarization vectors are
\begin{equation}\label{eq:polarization2D}
  e^1_k=(\cos\theta_k,\sin\theta_k)^T,\qquad e^2_k=(-\sin\theta_k,\cos\theta_k),
\end{equation}where $\theta_k$ parameterizes the direction of $k$.

It is not hard to check that the following matrix $M_0$ symplectically diagonalizes $H_0$ as in \eqref{eq:SymplecticDiag1}:
\begin{equation}\label{}
  M_0(k)=\frac{1}{\sqrt{2}}
\begin{pmatrix}
  I_2   & i I_2 \\
  I_2 & -i I_2
\end{pmatrix}
\begin{pmatrix}
  \begin{pmatrix}
    \sqrt{\rho E^1(k)} (e_k^1)^T \\
    \sqrt{\rho E^2(k)} (e_k^2)^T
  \end{pmatrix} & 0 \\
  0 & \begin{pmatrix}
        \frac{1}{\sqrt{\rho E^1(k)}} (e_k^1)^T \\
        \frac{1}{\sqrt{\rho E^2(k)}} (e_k^2)^T
      \end{pmatrix}
\end{pmatrix},
\end{equation} where $I_2$ denotes 2 by 2 identity matrix.

Applying first order perturbation theory in $\eta$, we get
\begin{equation}\label{}
  M_1(k)=\frac{i k \eta}{4\rho\sqrt{w_1w_2}}
\left(
\begin{array}{cccc}
 0 & \frac{w_1+w_2}{w_1-w_2} & 0 & \frac{w_1-w_2}{w_1+w_2} \\
 \frac{w_1+w_2}{w_1-w_2} & 0 & \frac{w_1-w_2}{w_1+w_2} & 0 \\
 0 & \frac{w_2-w_1}{w_1+w_2} & 0 & \frac{-w_1-w_2}{w_1-w_2} \\
 \frac{w_2-w_1}{w_1+w_2} & 0 & \frac{-w_1-w_2}{w_1-w_2} & 0 \\
\end{array}
\right).
\end{equation}

\subsection{Scattering rates}

  The disorder potential is given by \eqref{eq:Vdis}, with a $Q$-matrix representation as in \eqref{eq:Qdef} and
\begin{equation}\label{eq:Q2}
  Q_{JI}(q,p)=\begin{pmatrix}
                a\vec{p}\cdot \vec{q}+b \vec{p}^2\vec{q}^2 & 0 \\
                0 & 0
              \end{pmatrix}\otimes I_2.
\end{equation}

  Converting the $Q$-matrix into $F$-matrix as in Sec.~\ref{sec:Vsp}, we obtain
\begin{equation}\label{}
  F_{\alpha\beta}(p,q)=F^{(0)}_{\alpha\beta}(p,q)+F^{(1)}_{\alpha\beta}(p,q),
\end{equation}where
\begin{equation}\label{}
  F^{(0)}_{\alpha\beta}(p,q)=
\left(
\begin{array}{cc}
 \frac{\sqrt{p q} \cos \left(\theta _{p q}\right) \left(a \cos \left(\theta _{p q}\right)+b p q\right)}{\rho  w_1} & \frac{\sqrt{\frac{p q}{w_1 w_2}} \sin \left(\theta _{p q}\right) \left(a \cos \left(\theta _{p q}\right)+b p q\right)}{\rho } \\
 -\frac{\sqrt{\frac{p q}{w_1 w_2}} \sin \left(\theta _{p q}\right) \left(a \cos \left(\theta _{p q}\right)+b p q\right)}{\rho } & \frac{\sqrt{p q} \cos \left(\theta _{p q}\right) \left(a \cos \left(\theta _{p q}\right)+b p q\right)}{\rho  w_2} \\
\end{array}
\right),
\end{equation}
\begin{equation}\label{}
  F^{(1)}_{\alpha\beta}(p,q)=\eta
\left(
\begin{array}{cc}
 -\frac{i \sqrt{p q} (p+q) \sin \left(\theta _{p q}\right) \left(a \cos \left(\theta _{p q}\right)+b p q\right)}{\rho ^2 \left(w_1^2-w_2^2\right)} & \frac{i \sqrt{\frac{p q}{w_1 w_2}} \left(p w_1-q w_2\right) \cos \left(\theta _{p q}\right) \left(a \cos \left(\theta
   _{p q}\right)+b p q\right)}{\rho ^2 \left(w_1^2-w_2^2\right)} \\
 \frac{i \sqrt{\frac{p q}{w_1 w_2}} \left(p w_2-q w_1\right) \cos \left(\theta _{p q}\right) \left(a \cos \left(\theta _{p q}\right)+b p q\right)}{\rho ^2 \left(w_1^2-w_2^2\right)} & \frac{i \sqrt{p q} (p+q) \sin \left(\theta _{p q}\right) \left(a \cos \left(\theta
   _{p q}\right)+b p q\right)}{\rho ^2 \left(w_1^2-w_2^2\right)} \\
\end{array}
\right).
\end{equation} Here $\theta_{pq}=\theta_p-\theta_q$ is the angle between $p,q$. The above expressions can be further simplified by noticing that energy is conserved during collisions, so we can rewrite $p,q$ in terms of the conserved energy $E$, which yields
\begin{equation}\label{}
 F^{(0)}_{\alpha\beta}(p,q)=
\left(
\begin{array}{cc}
 \frac{E \cos \left(\theta _{p q}\right) \left(a w_1^2 \cos \left(\theta _{p q}\right)+b E^2\right)}{\rho  w_1^4} & \frac{E \sin \left(\theta _{p q}\right) \left(a w_1 w_2 \cos \left(\theta _{p q}\right)+b E^2\right)}{\rho  w_1^2 w_2^2} \\
 -\frac{E \sin \left(\theta _{p q}\right) \left(a w_1 w_2 \cos \left(\theta _{p q}\right)+b E^2\right)}{\rho  w_1^2 w_2^2} & \frac{E \cos \left(\theta _{p q}\right) \left(a w_2^2 \cos \left(\theta _{p q}\right)+b E^2\right)}{\rho  w_2^4} \\
\end{array}
\right),
\end{equation}
\begin{equation}\label{}
   F^{(1)}_{\alpha\beta}(p,q)=\eta
\left(
\begin{array}{cc}
 -\frac{2 i E^2 \sin \left(\theta _{p q}\right) \left(a w_1^2 \cos \left(\theta _{p q}\right)+b E^2\right)}{\rho ^2 w_1^4 \left(w_1^2-w_2^2\right)} & 0 \\
 0 & -\frac{2 i E^2 \sin \left(\theta _{p q}\right) \left(a w_2^2 \cos \left(\theta _{p q}\right)+b E^2\right)}{\rho ^2 w_2^4 \left(w_2^2-w_1^2\right)} \\
\end{array}
\right).
\end{equation}

We can then obtain the symmetric scattering rate and antisymmetric scattering rates as
\begin{equation}\label{eq:gammaSval}
  \gamma^S_{\alpha\beta}(p,q)=\frac{2\pi n_i}{\rho^2 V_s}
\left(
\begin{array}{cc}
 \frac{\cos ^2\left(\theta _{p q}\right) \left(a w_1^2 \cos \left(\theta _{p q}\right)+b E^2\right){}^2}{w_1^8} & \frac{\sin ^2\left(\theta _{p q}\right) \left(a w_1 w_2 \cos \left(\theta _{p q}\right)+b E^2\right){}^2}{w_1^4 w_2^4} \\
 \frac{\sin ^2\left(\theta _{p q}\right) \left(a w_1 w_2 \cos \left(\theta _{p q}\right)+b E^2\right){}^2}{w_1^4 w_2^4} & \frac{\cos ^2\left(\theta _{p q}\right) \left(a w_2^2 \cos \left(\theta _{p q}\right)+b E^2\right){}^2}{w_2^8} \\
\end{array}
\right),
\end{equation}
and
\begin{equation}\label{eq:gammaAval}
  \gamma^{A}_{\alpha\beta}(p,q)=\frac{\pi  E^5 \eta  n_i \sin (\theta _{pq})}{2 \rho ^4 w_1^6 w_2^6 V_s} \Gamma_{\alpha\beta},
\end{equation} with
\begin{equation}\label{}
  \Gamma_{11}=-\frac{\left(a w_1^2 \cos \left(\theta _{p q}\right)+b E^2\right) \left(a^2 w_1^4 w_2^2 \left(w_1^2+w_2^2\right) \cos \left(2 \theta _{p q}\right)+4 b^2 E^4 \left(w_1^4+w_2^2 w_1^2+w_2^4\right) \cos \left(\theta _{p q}\right)\right)}{w_1^8},
\end{equation}
\begin{equation}\label{}
\begin{split}
  \Gamma_{12}=\Gamma_{21}&=-\frac{1}{2} a^3 \left(w_1^2+w_2^2\right) \left(\cos \left(\theta _{p q}\right)+\cos \left(3 \theta _{p q}\right)\right)
-\frac{a^2 b E^2 \left(w_1^2+w_2^2\right) \cos \left(2 \theta _{p q}\right)}{w_1 w_2}\\
&-\frac{4 a b^2 E^4 \left(w_1^4+w_2^2 w_1^2+w_2^4\right) \cos
   ^2\left(\theta _{p q}\right)}{w_1^3 w_2^3}-\frac{4 b^3 E^6 \left(w_1^4+w_2^2 w_1^2+w_2^4\right) \cos \left(\theta _{p q}\right)}{w_1^4 w_2^4},
\end{split}
\end{equation}
\begin{equation}\label{}
  \Gamma_{22}=
-\frac{\left(a w_2^2 \cos \left(\theta _{p q}\right)+b E^2\right) \left(a^2 w_1^2 w_2^4 \left(w_1^2+w_2^2\right) \cos \left(2 \theta _{p q}\right)+4 b^2 E^4 \left(w_1^4+w_2^2 w_1^2+w_2^4\right) \cos \left(\theta _{p q}\right)\right)}{w_2^8}.
\end{equation}

\subsection{Solving the Boltzmann Equation}

  In two dimension, we can solve the Boltzmann equation analytically by generalizing the methods in \cite{SL03,Sinova07}. The Boltzmann equation takes the form
  \begin{equation}\label{eq:BE2}
  -\frac{\partial n_B}{\partial E_l}\frac{E_l}{T}|\nabla T||\vec{v}_l|\cos\phi_l=-I[f_l].
\end{equation}  We remind the reader that the phonon state label $l$ contains momentum and band index. Here $\phi_l$ is the angle between the velocity $\vec{v}_l$ and $\nabla T$.

  We can consider an ansatz of the form
\begin{equation}\label{eq:tauansatz}
  f_l=-\frac{\partial n_B}{\partial E_l}\left(-\frac{E_l |\nabla T|}{T}\right)|\vec{v}_l|\left[\tau_l^S \cos\phi_l +\tau_l^A\sin\phi_l\right],
\end{equation} where $\tau_l^S$ and $\tau_l^A$ are coefficients that \emph{only} depends on the band index of state $l$. They can be physically interpreted as relaxation times. Following calculations in Appendix.~\ref{sec:Boltz2D}, we obtain $\tau_l^S$ and $\tau_l^A$ to linear order in $\eta$ as (we have chosen $\nabla T$ to be along $\hat{x}$ direction so that $\phi_l$ coincides with $\theta_l=\theta_k$)
\begin{equation}\label{eq:taus}
\begin{split}
 \tau^{S}_{\alpha}&=\frac{8\rho^2 w_\alpha^{10}w_{\bar{\alpha}}^{6}}{E^3 n_i}\frac{\cN_\alpha^S}{\cD^S}\,,\\
 \cN_\alpha^S&=a^2 w_{\alpha }^2 w_{\bar{\alpha }}^4 \left(w_{\bar{\alpha }}^4+3 w_{\alpha }^4\right)+2 a b E^2 w_{\alpha }^2 w_{\bar{\alpha }}^2 \left(w_{\bar{\alpha }}^4-3 w_{\alpha }^4\right)+4 b^2 E^4 \left(w_{\bar{\alpha }}^6+w_{\alpha }^6\right)\,, \\
 \cD^S &=a^4 w_{\alpha }^6 w_{\bar{\alpha }}^6 \left(3 w_{\alpha }^8+10 w_{\alpha }^4 w_{\bar{\alpha }}^4+3 w_{\bar{\alpha }}^8\right)-6 a^3 b E^2 w_{\alpha }^4 w_{\bar{\alpha }}^4 \left(w_{\alpha }^{10}+3 w_{\alpha }^6 w_{\bar{\alpha }}^4+3 w_{\alpha }^4 w_{\bar{\alpha
   }}^6+w_{\bar{\alpha }}^{10}\right)\\
   &+4 a^2 b^2 E^4 w_{\alpha }^2 w_{\bar{\alpha }}^2 \left(w_{\alpha }^{12}+3 w_{\alpha }^{10} w_{\bar{\alpha }}^2+3 w_{\alpha }^8 w_{\bar{\alpha }}^4+10 w_{\alpha }^6 w_{\bar{\alpha }}^6+3 w_{\alpha }^4 w_{\bar{\alpha }}^8+3
   w_{\alpha }^2 w_{\bar{\alpha }}^{10}+w_{\bar{\alpha }}^{12}\right)\\
   &-24 a b^3 E^6 w_{\alpha }^2 w_{\bar{\alpha }}^2 \left(w_{\alpha }^{10}+w_{\alpha }^6 w_{\bar{\alpha }}^4+w_{\alpha }^4 w_{\bar{\alpha }}^6+w_{\bar{\alpha }}^{10}\right)+16 b^4 E^8 \left(w_{\alpha
   }^6+w_{\bar{\alpha }}^6\right)^2\,, \\
 \tau^A_\alpha & = \frac{4 a b \eta E^2 w_\alpha^{6}w_{\bar{\alpha}}^{6}}{n_i} \frac{\cN_\alpha^A}{\cD^A}\,, \\
      \cN_\alpha^A & =\left(a w_{\alpha }^2 w_{\bar{\alpha }}^2 \left(w_{\bar{\alpha }}^2+w_{\alpha }^2\right)-2 b E^2 \left(w_{\alpha }^2 w_{\bar{\alpha }}^2+w_{\bar{\alpha }}^4+w_{\alpha }^4\right)\right) \\
      &\times\left(a^2 w_{\alpha }^2 w_{\bar{\alpha }}^4 \left(w_{\bar{\alpha }}^4+3 w_{\alpha
   }^4\right)-4 a b E^2 w_{\alpha }^6 w_{\bar{\alpha }}^2+4 b^2 E^4 \left(w_{\bar{\alpha }}^6+w_{\alpha }^6\right)\right)\\
    &\times\left(a^2 w_{\alpha }^2 w_{\bar{\alpha }}^2 \left(w_{\bar{\alpha }}^2+w_{\alpha }^2\right){}^3-4 a b E^2 w_{\alpha }^2 w_{\bar{\alpha }}^2
   \left(w_{\bar{\alpha }}^4+w_{\alpha }^4\right)+8 b^2 E^4 \left(w_{\bar{\alpha }}^6+w_{\alpha }^6\right)\right)\,,\\
   \cD^A &= (\cD^S)^2\,.
  \end{split}
\end{equation}Here $\alpha=1,2$ is the band index, and $\bar{\alpha}=3-\alpha$.

We can proceed to compute the thermal conductivities using
\begin{eqnarray}
  \kappa_{xx} &=& \frac{1}{V_s}\sum_{l}\left(-\frac{\partial n_B}{\partial E_l}\right)\frac{E_l^2}{T}|\vec{v}_l|^2\cos^2\theta_l \tau_{l}^S, \\
  \kappa_{yx} &=& \frac{1}{V_s}\sum_{l}\left(-\frac{\partial n_B}{\partial E_l}\right)\frac{E_l^2}{T}|\vec{v}_l|^2\sin^2\theta_l \tau_{l}^A.
\end{eqnarray} The results are
\begin{equation}\label{eq:kappaxx2D}
  \kappa_{xx}=\frac{2 \rho ^2 w_1^6 w_2^6 }{\pi  T n_i}\int_0^{\infty}\rd E \left(-\frac{\partial n_B}{\partial E}\right)\frac{w_1^4 \cN^S_1+w_2^4 \cN^S_2}{\cD^S}\,,
\end{equation}
\begin{equation}\label{eq:kappayx2D}
  \kappa_{yx}=\frac{a b \eta w_1^6 w_2^6}{\pi T n_i}\int_0^{\infty}\rd E \left(-\frac{\partial n_B}{\partial E}\right) E^5 \frac{\cN^A_1+\cN^A_2}{\cD^A}\,.
\end{equation}

 The qualitative features of the above results agree with our general analysis in Sec.~\ref{sec:general}:
\begin{enumerate}
  \item Both $\kappa_{xx}$ and $\kappa_{xy}$ are proportional to $1/n_i$, i.e. proportional to mean-free-path. Therefore the heat conduction is enhanced in clean samples.
  \item The thermal Hall conductivity $\kappa_{xy}\propto ab$, therefore we need both scattering channels in \eqref{eq:Vdis} to produce non-zero thermal Hall effect. This agrees with the general analysis based on parity symmetry. This result continues to hold if the crystal is not isotropic but still has parity symmetry. The effects of introducing such anisotropicity are the following: a) The polarizations will not be characterized by $\theta_k$ but another angle $\vartheta_k$. b) The equal energy surface will not be circular, so $w_1,w_2$ become functions of $\theta_k$. c) The velocity is not parallel to momentum anymore, so we can't replace $\phi_k$ by $\theta_k$. However, all new functions introduced above only corrects $\gamma^A_{ll'}$ from the isotropic result by even harmonics in $\theta_k$, but from \eqref{eq:Mparallelmu} and \eqref{eq:Mperpmu} we need odd harmonics to have nonzero $\tau^A$, so we would still need two scattering channels of different parity.
  \item The longitudinal thermal conductivity $\kappa_{xx}$ has an $1/E$ divergence in the IR, which we naively regulate some cutoff $\Delta$ in the integral \eqref{eq:kappaxx2D}. As we shall see in later sections the more correct treatment is to consider boundary effects. At high temperature $\kappa_{xx}$ approaches a constant.
  \item The thermal Hall conductivity scales as $T^4$ at low temperature. At high temperature $T\gg T_{imp}$ it approaches a constant. The detailed behavior of $\kappa_{xy}$ in the crossover regime depends on microscopic details of the system. For instance, depending on the values of impurity couplings $a,~b$, $\kappa_{xy}$ might change sign as temperature rises.

\end{enumerate}

\begin{figure}
  \centering
  \begin{subfigure}{0.9\textwidth}
   \centering
   \includegraphics[width=\textwidth]{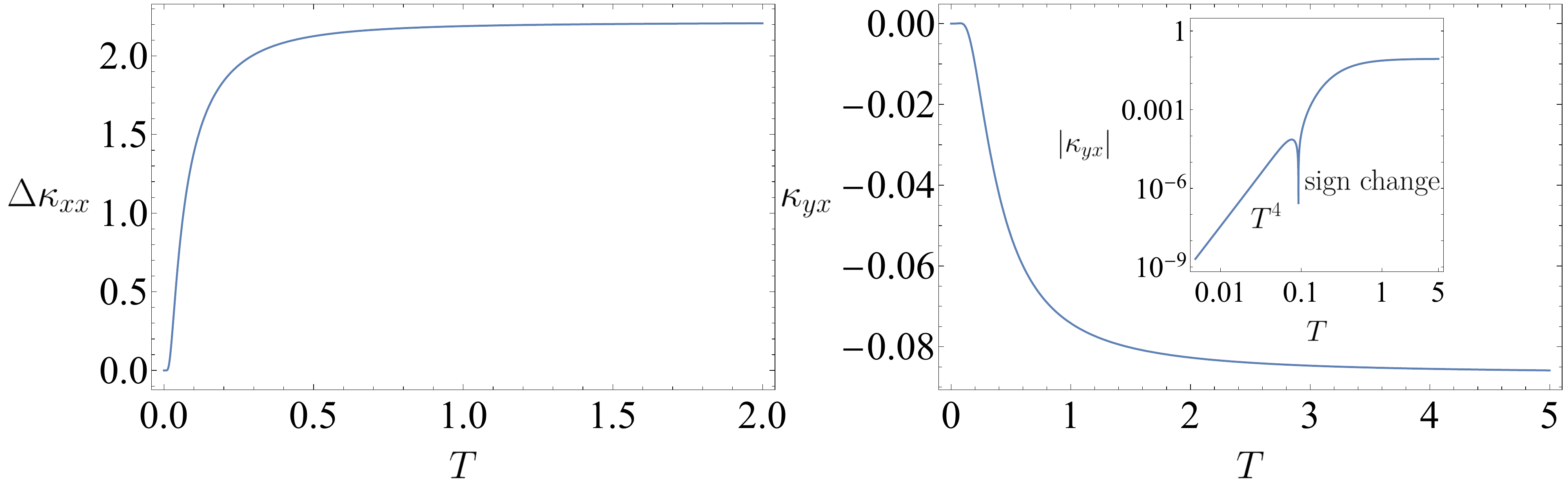}
  \end{subfigure}
  \begin{subfigure}{0.5\textwidth}
   \centering
   \includegraphics[width=\textwidth]{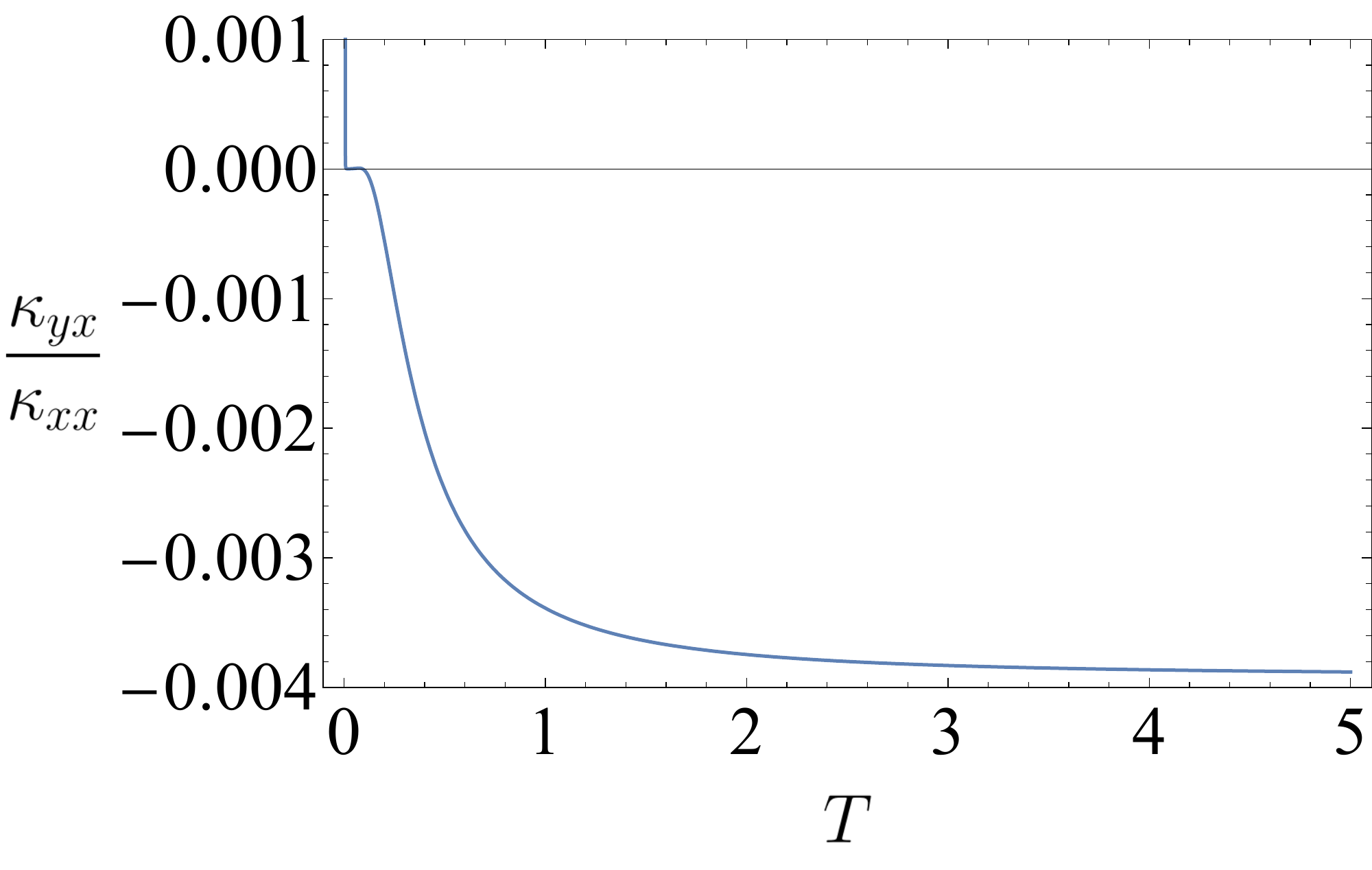}
  \end{subfigure}
  \caption{Left: $\Delta\kappa_{xx}(T)$; Right: $\kappa_{yx}(T)$, and the inset highlights some features in log-log scale; Bottom: The ratio $\kappa_{yx}/\kappa_{xx}$. Parameters used: $a=b=n_i=\rho=w_2=1,w_1=2$. $\kappa_{xx}$ is evaluated with a low energy cutoff $\Delta=0.1$ in \eqref{eq:kappaxx2D}. $\kappa_{xx}$ decays exponentially below cutoff, and saturates to a constant when $T\gg T_{imp}$, where $T_{imp}\sim w\sqrt{|a/b|}$ is the crossover energy scale set by impurity couplings. The thermal hall conductivity $\kappa_{xy}$ scales as $T^4$ at low temperature and saturates to a constant at high temperature. The detailed behavior of $\kappa_{xy}$ in the crossover regime (e.g. sign changing) depends on microscopic details of the model, such as impurity couplings $a,b$. The ratio $\kappa_{yx}/\kappa_{xx}$ blows up at low temperature due to exponential decay of $\kappa_{xx}$ below cutoff scale $\Delta$, and saturates to a constant at high energy. }\label{fig:kappaplot2D}
\end{figure}

  A numerical plot of $\kappa_{xx}$ and $\kappa_{xy}$ is shown in Fig.~\ref{fig:kappaplot2D}

\section{Thermal Hall Effect in a 3D Tetragonal Crystal}
\label{sec:3D}

  In 3D, we have to calculate the thermal conductivities numerically. Although it's possible to analytically solve the model with an isotropic crystal, the two degenerate transverse bands of the isotropic crystal violates our assumption and is not practically relevant. The strategy is to compute $\cK_{xx}$ and $\cK_{yx}$ as defined in \eqref{eq:Kijfunc} on a discretized equal-energy surface. We discretize the equal-energy surface in momentum space with the Gauss-Legendre qudrature, and then follow steps in Secs.~\ref{sec:model},\ref{sec:scatter},\ref{sec:thermalHall} to compute the scattering rates and the matrix elements of the collision operators $I_S$ and $I_A$, and finally evaluate the inner product \eqref{eq:Kijfunc}. In practice we used $\sim 3000$ points on the to discretize the equal-energy surface. We discuss some details of inverting $I_S$ in Appendix.~\ref{sec:invertIS}.

\begin{figure}[htb]
  \centering
  \begin{subfigure}{0.4\textwidth}
    \centering
    \includegraphics[width=\textwidth]{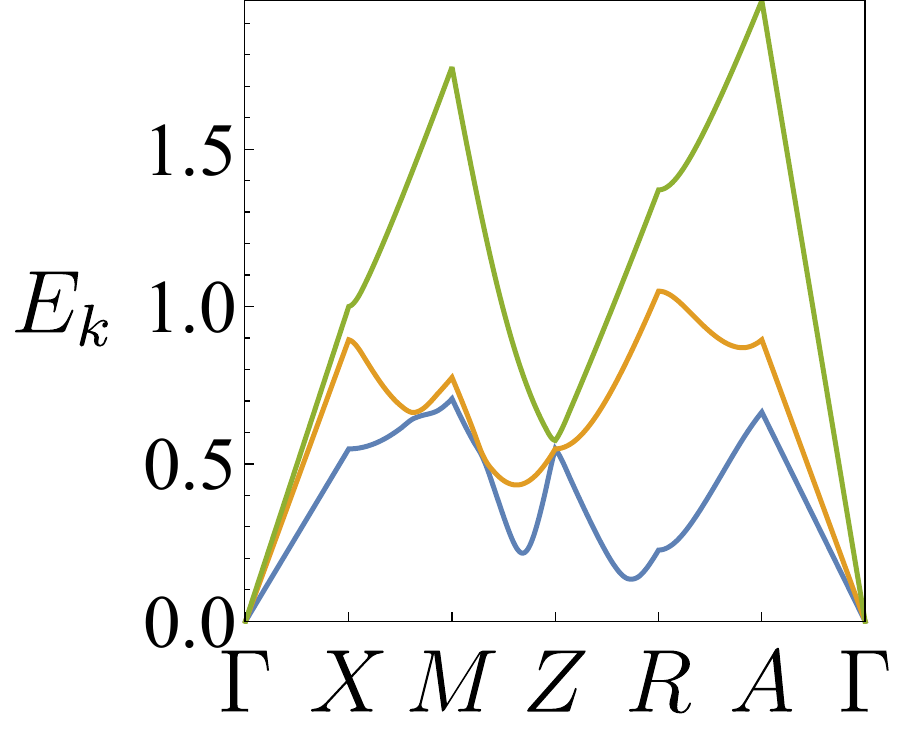}
  \end{subfigure}
   \begin{subfigure}{0.4\textwidth}
    \centering
    \includegraphics[width=\textwidth]{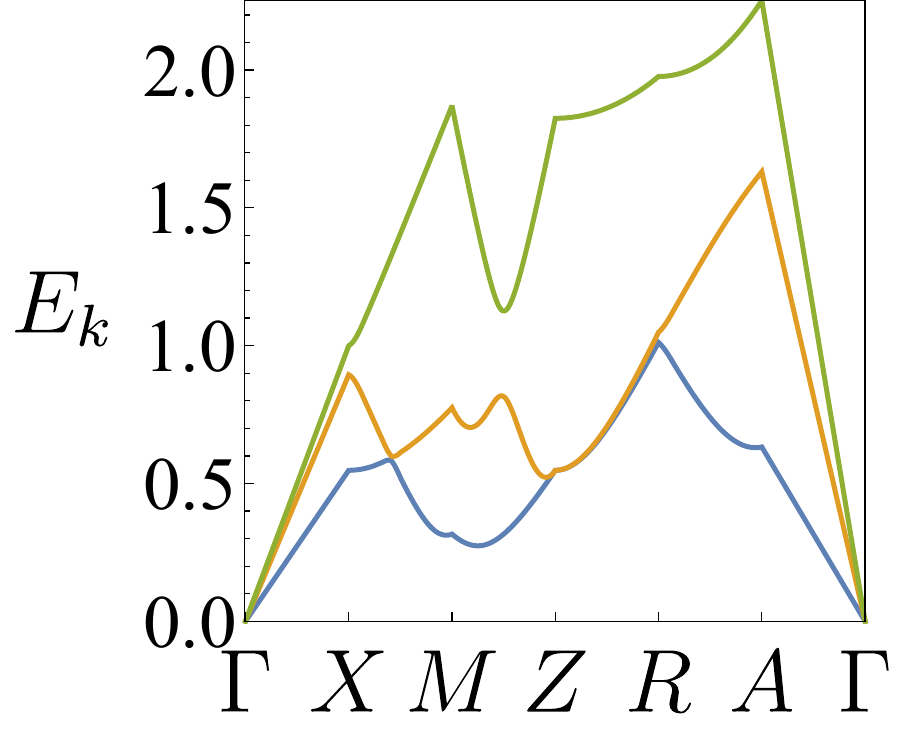}
  \end{subfigure}
  \caption{The two acoustic phonon band structures corresponding to two sets of parameters in Table.~\ref{tab:para}. The band structures are calculated from the elastic theory \eqref{eq:Lph}, where the phonon dispersion is linear. The phonon bands are non-degenerate except along the $k_z$ axis. \label{fig:bandplot}}
\end{figure}

  We consider two different sets of parameters as in Table.~\ref{tab:para}, whose band structures are shown in Fig.~\ref{fig:bandplot}.

\begin{table}
  \centering
  \begin{tabular}{|c|c|c|c|c|c|c|c|c|c|c|c|}
  \hline
  & $\eta$ & $\rho$ & $a$ & $b$ & $n_i$ & $C_{11}$ & $C_{12}$ & $C_{13}$ & $C_{44}$ & $C_{66}$ & $C_{33}$ \\
\hline
 (a) &\multirow{2}{2em}{1}&\multirow{2}{2em}{1} & \multirow{2}{2em}{1} & \multirow{2}{2em}{1}
 & \multirow{2}{2em}{1} & \multirow{2}{2em}{1} & 0.5 & \multirow{2}{2em}{0.55} & \multirow{2}{2em}{0.3} & \multirow{2}{2em}{0.8} & 0.33 \\
\cline{1-1} \cline{8-8} \cline{12-12}
 (b) &  &  &  &   &   &   & 0.9  &   &   &   & 3.33  \\
  \hline
\end{tabular}
  \caption{Parameters used for numerical calculation}\label{tab:para}
\end{table}

\begin{figure}[H]
  \centering
  \begin{subfigure}{0.9\textwidth}
   \centering
   \includegraphics[width=\textwidth]{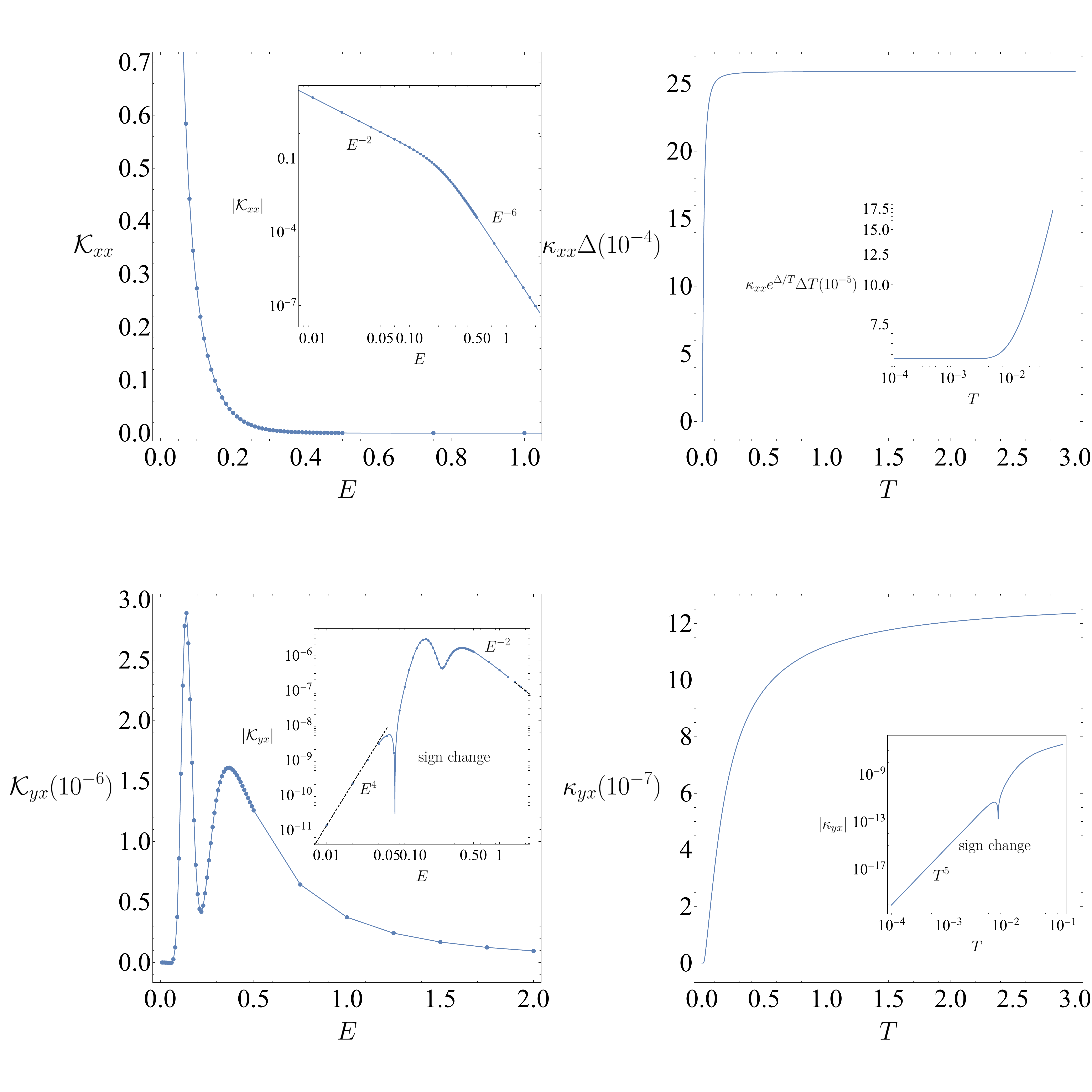}
  \end{subfigure}
  \caption{ The results of parameters (a). We have plotted $\cK_{xx}(E),~\kappa_{xx}(T),~\cK_{yx}(E)$, $\kappa_{yx}(T)$ . The insets are log-log scale plots where we highlight some of the features.
Here $\cK$ refers to spectral thermal conductivity and $\kappa$ refers to thermal conductivity, and they are related by Eq.~\eqref{eq:kappaK}. As we can see at low and high temperatures $\cK_{xx}$ and $\cK_{yx}$ scale as powers of energy $E$ with powers predicted in Sec.~\ref{sec:general}.
For the longitudinal thermal conductivity $\kappa_{xx}$, we have imposed an IR cutoff $\Delta=0.02$.  Below the cutoff, $\kappa_{xx}$ decays exponentially and saturates to a constant at high temperature. The thermal hall conductivity $\kappa_{yx}$ is proportional to $T^5$ at low temperatures and saturates to a constant at high temperature.
 All crossovers happen at the impurity scale $T_{imp}\sim w\sqrt{|a/b|}$. }\label{fig:result_a}
\end{figure}

\begin{figure}
  \centering
  \includegraphics[width=0.6\textwidth]{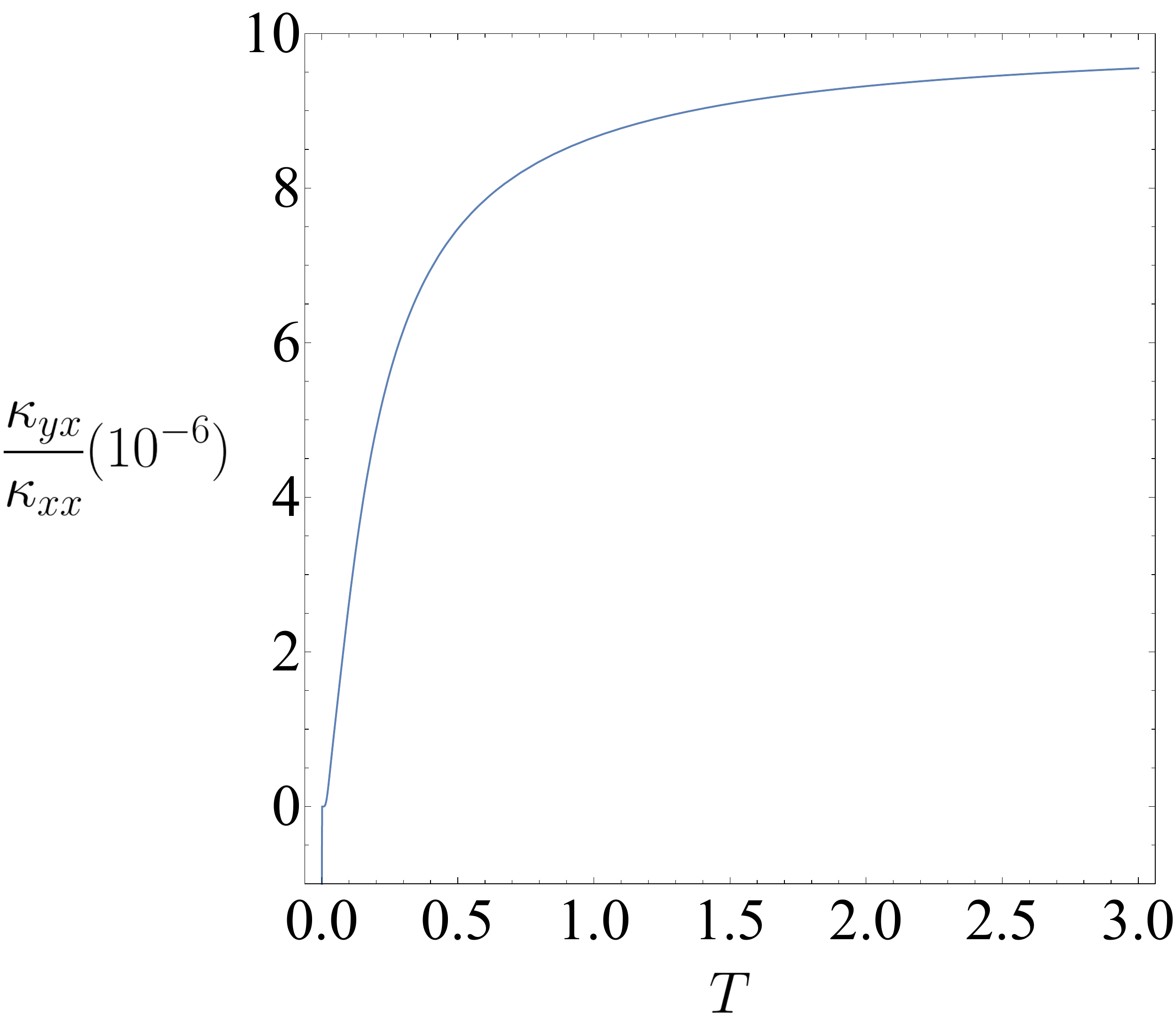}

  \caption{The results of parameter(a).  The ratio $\kappa_{yx}/\kappa_{xx}$ blows up at temperature below the cutoff due to exponential decay of $\kappa_{xx}$ and saturates to a constant when $T\gg T_{imp}$.}\label{fig:result_a2}
\end{figure}

\begin{figure}[H]
  \centering
  \begin{subfigure}{0.85\textwidth}
   \centering
   \includegraphics[width=\textwidth]{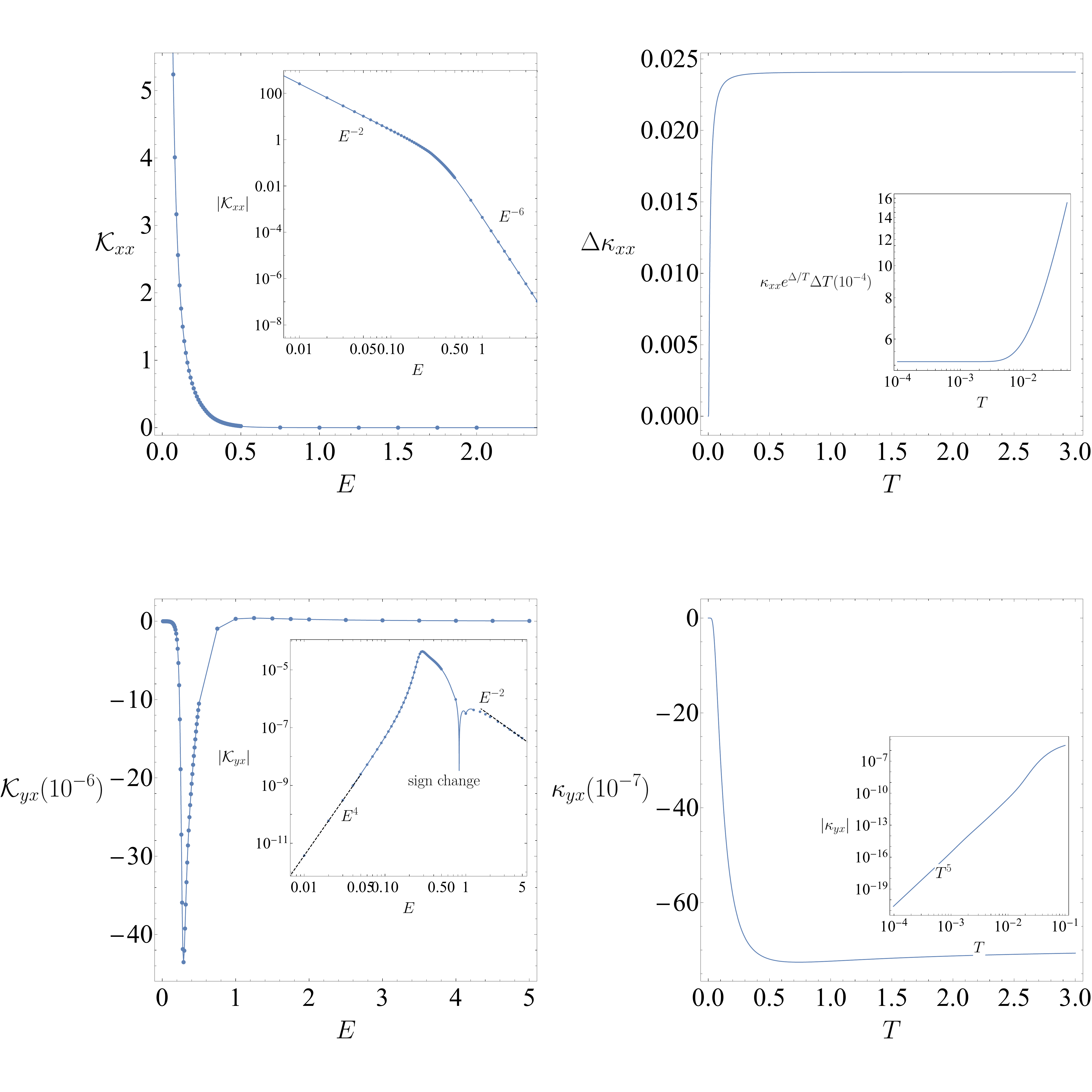}
  \end{subfigure}
  \begin{subfigure}{0.4\textwidth}
   \centering
   \includegraphics[width=\textwidth]{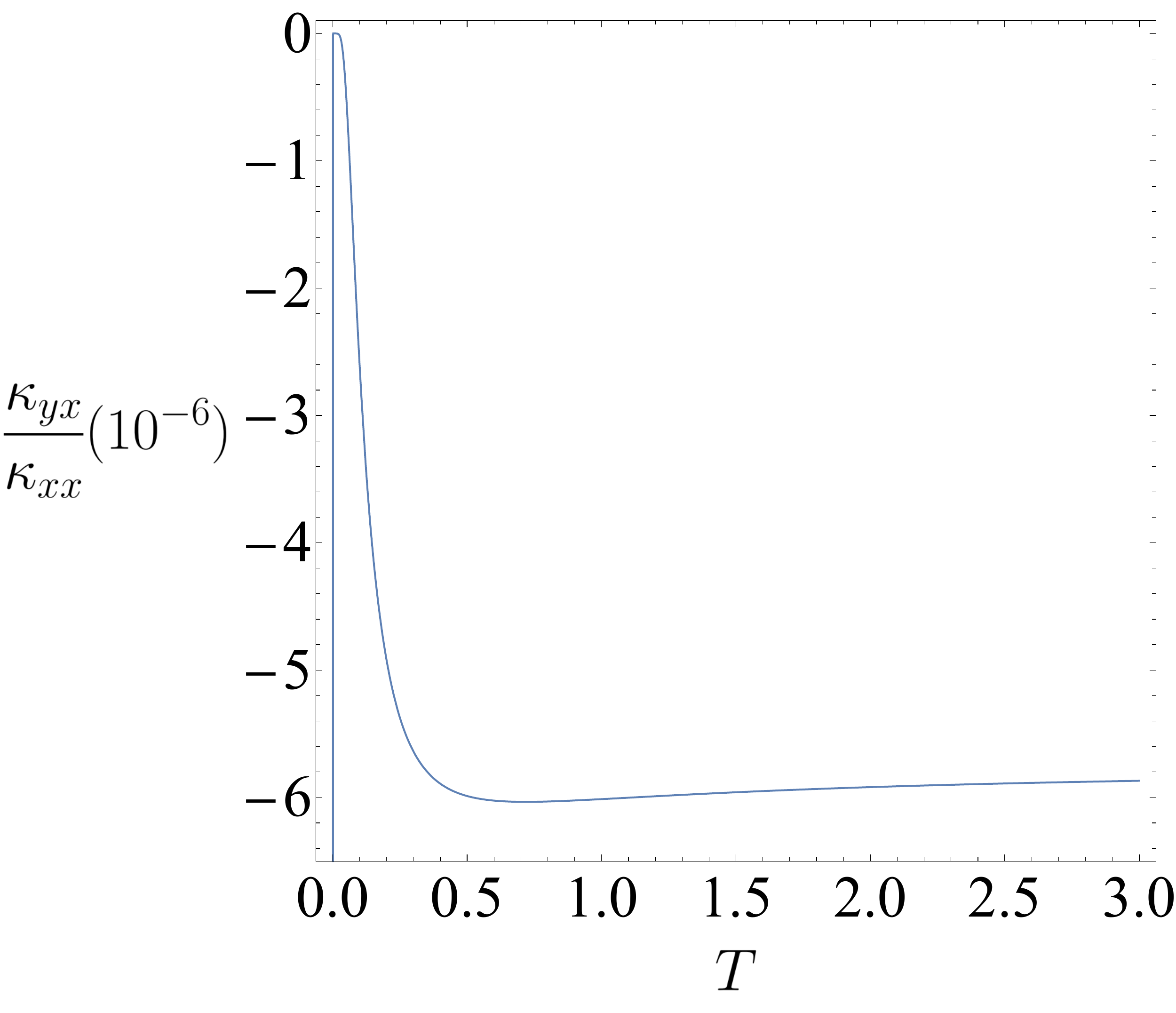}
  \end{subfigure}
  \caption{The results of parameters (b). Features similar to Fig.~\ref{fig:result_a} are also present here. The main difference is that $\kappa_{yx}$ has a different sign. This shows that the detailed behavior of $\kappa_{yx}$ in the crossover regime depends on microscopic details such as impurity couplings or band structure.}\label{fig:result_b}
\end{figure}

  The results are shown in Fig.~\ref{fig:result_a}, Fig.~\ref{fig:result_a2} and Fig.~\ref{fig:result_b}. In terms of scaling behavior, both the spectral thermal conductivity  $\cK_{ij}(E)$ and the thermal conductivities $\kappa_{ij}(T)$  agree with our general analysis in Sec.~\ref{sec:general}, and are similar to the features seen in the 2D calculation in Sec.~\ref{sec:2D}. Comparing the two sets of parameters (a) and (b), we conclude that some of the features such as peaks and signs in $\cK_{yx}$ depend on details of the phonon band structure and phonon-disorder interaction.


\section{Boundary Scattering in 2D}
\label{sec:boundary}

\begin{figure}
  \centering
  \includegraphics[width=0.6\textwidth]{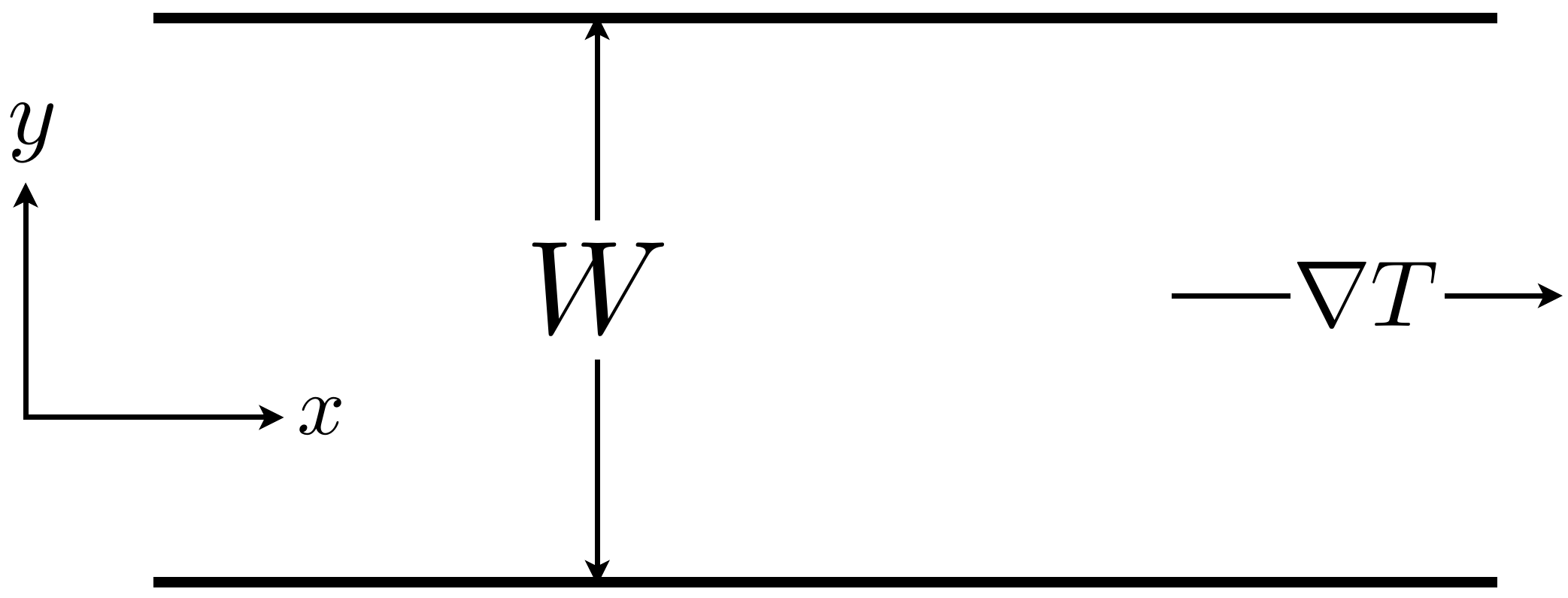}
  \caption{Thermal transport in a semi-infinite sample of width $w$. In general, there is scattering of phonons off impurities in the sample, and non-specular reflection off the boundaries. The thermal Hall transport arises from the phonon Hall viscosity, which is assumed to be present in the bulk.\label{fig:geometry}}
\end{figure}

The IR divergence of the longitudinal thermal conductivity means that at low energy the primary scattering mechanism is not disorder, but boundaries of the sample. In our analysis so far, we simply accounted for this by introducing the cutoff energy $\Delta$. In this section we present an analysis of the Boltzmann equation in a slab geometry, where the slab width $W$ will serve as an IR cutoff. We assume the slab has infinite length. To make the problem analytically tractable, we will only consider the isotropic 2D crystal. We will be focusing on the low temperature limit where $W$ is small compared to phonon mean-free path. See Fig.~\ref{fig:geometry} for the geometry of the thermal transport.

\subsection{Boltzmann equation in presence of boundary}
  The full Boltzmann equation is
\begin{equation}\label{}
  \partial_t f_l+\vec{v}_l\cdot \nabla f_l=-I[f_l].
\end{equation} Here the collision integral $I=I_S+I_A+I_B$ includes symmetric scattering due to impurities, skew-scattering due to impurities and boundary scattering. We assume they all conserve phonon number and energy. Assuming the distribution function takes the form $f_l=n_B(E_l-\phi_l)$, and linearize around equilibrium, we obtain
\begin{equation}\label{}
  \partial_t \phi_l+\vec{v}_l\cdot (\nabla \phi_l+\frac{E_l\nabla T}{T})=-I[\phi_l].
\end{equation} Since we have assumed energy conservation, $\phi_l$'s of different energy surfaces are decoupled from each other, and we can focus on solving the equation on the $E_l=E$ surface. 

  Before diving into detailed analysis, we make some intuitive discussion. Due to the decoupling of states of different energy, the thermal conductivity is some weighted integral of conductivities on different equal-energy surfaces. For states of high energy, the bulk scattering is dominant and we should have the usual bulk transport behavior. For states of low energy, the bulk relaxation time diverges as $1/E^3$, and the boundary contribution is dominant. Note that the transport problem in this case is very similar to that of ballistic electrons, so we expect to get a thermal conductivity described by Fuchs-Sondheimer regime of transport \cite{Sondheimer2001,AS2018,AS2019}. The mean free-path will scale as $l\sim W \ln(w\tau/W)$, where $w$ is phonon velocity, $\tau$ is the bulk relaxation time and $W$ is the slab width. The logarithmic enhancement factor is due to particles travelling almost parallel to the slab.

Our analysis generalizes \cite{Sondheimer2001,AS2018,AS2019} to the two-band case. In the steady state, the distribution function $\phi_l$ has the form $\phi_l=\phi_{\alpha}(\vec{x},\theta)$, where $\alpha=1,2$ is the phonon band index, $\vec{x}$ is real-space coordinate and $\theta$ is the angular direction of phonon momentum. We will also use the angular harmonics in $\theta$, denoted by
\begin{equation}\label{}
  \phi_{\alpha}^m(\vec{x})=\int_0^{2\pi}\frac{\rd \theta}{2\pi} e^{-im\theta}\phi_{\alpha}(\vec{x},\theta).
\end{equation} Since we are assuming an isotropic 2D crystal, the zeroth harmonics in $\theta$ is associated with phonon number density and the first harmonics in $\theta$ is associated with phonon current.

We take the slab geometry to be described by coordinate $(x,y)$ where $x\in \mathbb{R}$ and $y\in[0,W]$. We assume the boundary scatterings at $y=0$ and $y=W$ are completely diffusive: All incoming particles are reflected to each direction and each band with equal probability. The boundary condition can be written as
\begin{equation}\label{eq:BC}
  \phi_\alpha(y=0,0<\theta<\pi)=c_D[\phi]\,,\qquad \phi_\alpha(y=W,-\pi<\theta<0)=c_U[\phi]\,,
\end{equation}where
\begin{equation}\label{eq:cDcU}
\begin{split}
  c_D[\phi]&=-\frac{1}{4}\sum_{\beta}\int_{-\pi}^{0} \rd \theta \phi_\beta(y=0,\theta)\sin\theta\,,\\
  c_U[\phi]&=\frac{1}{4}\sum_{\beta}\int_{0}^{\pi} \rd \theta \phi_\beta(y=W,\theta)\sin\theta\,.
\end{split}
\end{equation} In the literature, our boundary condition is referred as non-thermalizing \cite{RM2016}, as opposed to thermalizing boundary conditions where $\phi_{\alpha}$ is determined solely by boundary temperature. It is shown in Ref.~\cite{RM2016} that the two boundary conditions yield the same transport coefficients in the steady state and the non-thermalizing boundary condition is more physical in the transient state because it respects heat flux conservation.

We discuss possible generalization of our diffusive boundary condition. In the Fuchs-Sondheimer theory \cite{Sondheimer2001,RM2016}, the diffusive boundary condition can be generalized to a mixture of diffusive + specular boundary condition, and this renormalizes the conductivity by a function of the specularity parameter $p$, the portion of specular scattering. For realistic materials, $p$ is not very close to one, and the renormalization factor is of order one. We also expect that we can make the probability of scattering to each band different and maintain the essential physics, so the current assumption of equal probability in each band is for convenience of analysis. We have also assumed that the scattering at the boundary respects time-reversal symmetry. Time-reversal breaking scattering at the boundary might also contribute to the thermal hall transport, but it's not yet clear how to describe in the Boltzmann formalism and we leave it for future study.

The collision operator $I[\phi_l]$ can be decomposed into two parts $I=I_S+I_A$ meaning normal (symmetric) scattering and skew (antisymmetric) scattering. Due to rotation symmetry, the collision integral preserves angular harmonics.

 Using the definition of the collision integral \eqref{eq:If}, we can write down the matrix element of the collision operator $I_S$:
\begin{equation}\label{}
  I_S=\hat{\Gamma}-\hat{\gamma}^S.
\end{equation} Here $\hat{\Gamma}$ corresponds to the first term in \eqref{eq:If} (the departure term), which is diagonal in both harmonics and band indices
\begin{equation}\label{}
  \hat{\Gamma}[\phi]^n_\alpha(\vec{x})=\hat{\Gamma}_{\alpha}\phi^n_\alpha(\vec{x}),
\end{equation} where
\begin{equation}\label{}
  \hat{\Gamma}_\alpha=V_s\sum_{\beta}\int \frac{E\rd \theta'}{w_\beta^2(2\pi)^2}\gamma^S_{\alpha\beta}(\theta,\theta'),
\end{equation} with $\gamma^S_{\alpha\beta}$ given by \eqref{eq:gammaSval}.

The operator $\hat{\gamma}^S$ corresponds to the second term (the arrival term) in \eqref{eq:If}. Because of rotational symmetry it is diagonal in harmonic index
\begin{equation}\label{}
  \hat{\gamma}^S[\phi]^n_\alpha(\vec{x})=\sum_{\beta} \hat{\gamma}^S_{n,\alpha\beta}\phi^n_\beta(\vec{x}),
\end{equation} where
\begin{equation}\label{eq:ghatSalphabeta}
  \hat{\gamma}^S_{n,\alpha\beta}=V_s \int \frac{E \rd \theta'}{w_\beta^2 (2\pi)^2} \gamma^S_{\alpha\beta}(\theta,\theta')e^{in(\theta'-\theta)}\,.
\end{equation}

The skew-scattering (antisymmetric) term $I_A$ only contains the arrival term
\begin{equation}\label{}
  I_A=-\hat{\gamma}^A\,,
\end{equation} where the matrix elements of $\hat{\gamma}^A$ are calculated similarly as \eqref{eq:ghatSalphabeta}.

The explicit values of $\hat{\Gamma}$, $\hat{\gamma}^S$ and $\hat{\gamma}^A$ are tabulated in Appendix.~\ref{sec:gammaval}.

\subsection{Solving the Boltzmann equation}

  The treatment of the problem is similar to the usual Fuchs-Sondheimer theory \cite{AS2018,AS2019,Sondheimer2001,RM2016}. We assume a temperature gradient in the $x$-direction, and that the distribution function only depends on $y$. The Boltzmann equation can be written as
\begin{equation}\label{eq:BEgamma}
 ( \hat{\Gamma}_\alpha +w_\alpha \sin \theta \partial_y)\phi_{\alpha}(y,\theta)-Xw_\alpha\cos\theta =(\hat{\gamma}^S+\hat{\gamma}^A)[\phi]_\alpha(\theta)\,,
\end{equation} where $X=-E \partial_x T/T$ is the driving force. In the ballistic limit, the RHS of \eqref{eq:BEgamma} is small and can be treated as a perturbation. The zeroth order solution $\phi^{(0)}_\alpha(y,\theta)$ therefore satisfies
\begin{equation}\label{}
   ( \hat{\Gamma}_\alpha +w_\alpha \sin \theta \partial_y)\phi^{(0)}_{\alpha}(y,\theta)-Xw_\alpha\cos\theta =0\,,
\end{equation}subject to boundary conditions \eqref{eq:BC}. After some manipulation, we found a solution satisfying zero boundary condition $c_D[\phi^{(0)}]=c_U[\phi^{(0)}]=0$ because of $\theta\rightarrow\pi-\theta$ symmetry, with
\begin{equation}\label{}
  \phi^{(0)}_\alpha(y,\theta)=X\frac{w_\alpha \cos\theta}{\hat{\Gamma}_\alpha}\begin{cases}
                                1-e^{-\frac{\hat{\Gamma}_{\alpha}y}{w_\alpha\sin\theta}}, & 0<\theta<\pi \\
                                1-e^{\frac{\hat{\Gamma}_{\alpha}(W-y)}{w_\alpha\sin\theta}}, & -\pi<\theta<0 .
                              \end{cases}
\end{equation}

We can compute the longitudinal thermal conductivity from the solution $\phi^{(0)}$. Using $f_l=n_B(E_l-\phi_l)$ and \eqref{eq:kappaK}, \eqref{eq:calK}, we can write down the spectral thermal conductivity as
\begin{equation}\label{}
  \cK_{xx}(E)=\int_0^W\frac{\rd y}{W}\int_0^{2\pi} \frac{\rd \theta}{2\pi} \sum_{\alpha}\frac{E}{2\pi w_\alpha^2} w_\alpha\cos\theta \frac{\phi_\alpha(y,\theta)}{X}\,.
\end{equation} Evaluating the integral, we get to leading log singularity that
\begin{equation}\label{}
  \cK_{xx}(E)=\sum_{\alpha}\frac{E W}{2\pi^2 w_{\alpha}}\ln\left(\frac{w_\alpha}{\hat{\Gamma}_\alpha W}\right)\,.
\end{equation}Here we are taking the limit $\hat{\Gamma}_\alpha W/w_\alpha\ll 1$ and only retained the leading logarithmic singularity. This gives rise an longitudinal thermal conductivity $\kappa_{xx}\sim T^2 W\ln(w/(WT))$ at low temperature. Note that this is the thermal conductivity in which the scattering is primarily from the boundary, and the dependence on impurity density only appears in the log factor via $\hat{\Gamma}$. The logarithmic factor reflects the fact that particles travelling almost parallel to the boundary have long mean free time and contributes the most to transport. The result can be generalized to spatial dimension $d$ with $\kappa_{xx}\sim T^d W \ln(w/(WT))$, which up to the log factor is the usual Casimir limit of boundary scattering thermal conductivity \cite{Casimir1938}.

The RHS of \eqref{eq:BEgamma} produces the so called hydrodynamic corrections to the solution $\phi^{(0)}$, which multiplies $\phi^{(0)}$ by some higher powers of $\hat{\Gamma}_{\alpha} W/w_{\alpha}\ll 1$. We will be only interested in those corrections that give rise to a thermal Hall effect, i.e. corrections due to $\hat{\gamma}^A$. Writing the solution as $\phi=\phi^{(0)}+\phi^{(1)}$, and expanding \eqref{eq:BEgamma} to first order in $\phi^{(1)}$ and $\hat{\gamma}^A$, we obtain
\begin{equation}\label{eq:BEphi1}
  (\hat{\Gamma}_\alpha+w_\alpha \sin\theta\partial_y)\phi_\alpha^{(1)}(y,\theta)=\hat{\gamma}^A[\phi^{(0)}]_\alpha(\theta)\,.
\end{equation}
To leading log singularity, the RHS of \eqref{eq:BEphi1} consists of four terms from four harmonic channels of $\hat{\gamma}^A$. The detailed calculation is in Appendix.~\ref{sec:BoltzBoundary} and the result is
\begin{equation}\label{eq:BEphi1RHS}
  \hat{\gamma}^A[\phi^{(0)}]_\alpha(\theta)=\frac{2}{\pi}X W(z_{s1,\alpha}\sin\theta+z_{s3,\alpha}\sin3\theta)+X(2y-W)(z_{c2,\alpha}\cos2\theta+z_{c4,\alpha}\cos 4\theta)\,,
\end{equation}where
\begin{equation}\label{}
\begin{split}
  z_{s1,\alpha}&=\sum_{\beta}i\hat{\gamma}^{A}_{+1,\alpha\beta}\ln\frac{w_\beta}{\hat{\Gamma}_\beta W}\,,\\
  z_{s3,\alpha}&=\sum_\beta i\hat{\gamma}^{A}_{+3,\alpha\beta}\ln\frac{w_\beta}{\hat{\Gamma}_\beta W} \,, \\
  z_{c2,\alpha}&=\sum_\beta -i\hat{\gamma}^A_{+2,\alpha\beta}\,, \\
  z_{c4,\alpha}&=\sum_\beta -i\hat{\gamma}^A_{+4,\alpha\beta}\,.
\end{split}
\end{equation}

The solution of \eqref{eq:BEphi1} consists of four terms contributed from each of the scattering channels. The detailed solution is written in Appendix.~\ref{sec:BoltzBoundary}.

To extract the thermal Hall conductivity, we need to calculate the Hall temperature gradient which balances the Lorentz force acting on particles. According to Einstein relation, such a temperature gradient can be read off from the zeroth harmonics of $\phi_\alpha$, which we now calculate. It's easy to see that $\phi^{(0)}$ doesn't contain zeroth harmonics, and we just need to look at $\phi^{(1)}$. The difference of zeroth harmonics of $\phi^{(1)}_{\alpha}$ between $y=W$ and $y=0$ is given by
\begin{equation}\label{}
  \Delta \phi_{s1,\alpha,m=0}\vert_{y=0}^{y=W}=\frac{2 X W^2}{\pi}\frac{z_{s1,\alpha}}{w_\alpha}\,,
\end{equation}
\begin{equation}\label{}
  \Delta \phi_{s3,\alpha,m=0}\vert_{y=0}^{y=W}=\frac{2 X W^2}{\pi}\frac{z_{s3,\alpha}}{w_\alpha}\,,
\end{equation}
\begin{equation}\label{}
  \Delta \phi_{c2,\alpha,m=0}\vert_{y=0}^{y=W}\sim \Delta \phi_{c4,\alpha,m=0}\vert_{y=0}^{y=W}\sim X W^2 \frac{z_{c2/c4,\alpha}}{w_\alpha} \ln \frac{w_\alpha}{\hat{\Gamma}_\alpha W}\,.
\end{equation} Here we have presented $\Delta\phi$ for each of the harmonic channels in $\hat{\gamma}^A$, and the total $\Delta\phi$ is the sum of them. The last two results are only scale estimates, as the leading term in $\hat{\Gamma}_\alpha W/w_\alpha$ cancelled out.

From the above result we can roughly estimate that the effective Hall temperature gradient as
\begin{equation}\label{}
  X_H=\sum_{k,\alpha}p_\alpha\frac{\Delta \phi_{k,\alpha}\vert_{y=0}^{y=W}}{W}\sim \sum_{k,\alpha} p_\alpha X W \frac{z_{k,\alpha}}{w_\alpha}\,.
\end{equation} Here in the sum $k$ runs over $s1,s3,c2,c4$, and $p_\alpha=(\frac{w_2^2}{w_1^2+w_2^2},\frac{w_1^2}{w_1^2+w_2^2})$ is a projector that projects out the non-zero mode in the $m=0$ harmonic sector.
Using the relation of no transverse current $X_H \cK_{xx}+X \cK_{yx}=0$, we have
\begin{equation}\label{}
  \cK_{yx} =- \cK_{xx} \frac{X_H}{X}\sim \cK_{xx} W\frac{\gamma^A}{w}\ln\frac{w}{\hat{\Gamma} W}.
\end{equation} As an optimistic estimate, we assume a single scattering channel can contribute to thermal Hall. We take $\gamma^A$ to be the largest possible value $\gamma^A=\hat{\gamma}^A_{+4}\sim E^6$, and then $\cK_{yx}\sim E^7 W^2\ln^2 (E/T_b)$, which implies $\kappa_{yx}\sim T^8 W^2 \ln^2 (T/T_{b})$. Here $T_b$ is the energy/temperature scale where bulk mean-free path becomes comparable to slab width $W$, estimated from $\hat{\Gamma}/w\approx W$. We can see that even without considering two-channel scattering due to parity symmetry, the thermal hall effect due to boundary is smaller than the bulk result.

\section{Conclusions}
\label{sec:conc}

Our analysis has shown that computing the skew scattering contribution to phonon thermal Hall transport involves numerous subtleties that were not previously realized. We confirmed the $1/n_i$ scaling of the bulk thermal Hall conductivity as predicted by Chen \textit{et al.} \cite{kivelson20}, but the interplay between parity symmetry and phonon-impurity coupling shows the necessity to include multiple scattering channels, and this suppresses the impurity contribution to the  thermal Hall conductivity at the lowest temperatures, and the temperature scaling is $T^{d+2}$.  We note that our analysis considered impurities that were point-like {\it i.e.\/} smaller than the acoustic phonon wavelength.
Nevertheless, we do find a regime of temperature independent Hall conductivity at higher temperatures at a value which is sensitive to details of the phonon-impurity coupling and the phonon band structure. Given this sensitivity, quantitative general estimates are difficult to make. Nevertheless, we provide qualitative estimates for the different regimes of longitudinal and Hall transport in Section~\ref{sec:general}. We also carried out complete computations in simple models in 2 and 3 dimensions, and the results are in Figs.~\ref{fig:kappaplot2D}, \ref{fig:result_a} , \ref{fig:result_a2} and \ref{fig:result_b}.

We also considered the effect of non-specular scattering off sample boundaries in Section~\ref{sec:boundary}: although they do help to regularize the longitudinal thermal conductivity at lowest temperatures, they also further suppress the low temperature thermal Hall effect.

Comparing our result to the intrinsic thermal Hall conductivity in \cite{kivelson20,ZhangTeng21}, which scales as $T^d$, the skew scattering contribution is dominated by the intrinsic contribution as $T\to 0$. However, given the $1/n_i$ impurity enhancement, we do expect the skew scattering contribution to take over at some elevated temperature $T_*$ set by the impurity density. Therefore, our theory should be applicable in the regime $\max(T_*,T_b)<T<T_D$, where $T_D$ is the Debye temperature, and $T_b$ is the temperature scale where boundary effects become important.

\section*{Acknowledgements}
\label{sec:ack}

We thank Rhine Samajdar, Mathias Scheurer, Yanting Teng, and Yunchao Zhang for collaborations on related projects, and them and Jing-Yuan Chen, Ga{\"e}l Grissonnanche, Steven Kivelson, Xiao-Qi Sun, Leonid Levitov, and Loius Taillefer, for valuable discussions. This research was supported by the National Science Foundation under Grant No.~DMR-2002850. This work was also supported by the Simons Collaboration on Ultra-Quantum Matter, which is a grant from the Simons Foundation (651440, S.S.).

\appendix

\section{Solution of Boltzmann equation in 2D}\label{sec:Boltz2D}
 In this part we discuss the calculation of relaxation times defined in Eq.~\eqref{eq:tauansatz}.

  Acting the collision integral $I$ on the ansatz \eqref{eq:tauansatz}, we obtain
\begin{equation}\label{eq:Ifl}
  I[f_l]=-\frac{\partial n_B}{\partial E_l}\left(-\frac{E_l |\nabla T|}{T}\right)|\vec{v}_l|\left[(M_{l\parallel }^S+M_{l\perp }^A)\cos\phi_l+(M_{l\parallel}^A-M_{l\perp}^S)\sin\phi_l\right],
\end{equation} where
\begin{eqnarray}
  M_{l\parallel}^\mu &=& \sum_{l'}\gamma_{l'l}\tau_l^\mu-\gamma_{ll'}\frac{|\vec{v}_{l'}|}{|\vec{v}_l|}\cos(\phi_{l'}-\phi_{l})\tau_{l'}^{\mu} \,, \label{eq:Mparallelmu}\\
  M_{l\perp}^{\mu} &=& \sum_{l'}\gamma_{ll'}\frac{|\vec{v}_{l'}|}{|\vec{v}_l|}\sin(\phi_{l}-\phi_{l'})\tau_{l'}^{\mu}\,,\quad \mu=S,A\,.\label{eq:Mperpmu}
\end{eqnarray} Matching \eqref{eq:Ifl} with the LHS of \eqref{eq:BE2}, we obtain the following linear equations for $\tau^S_l$ and $\tau^A_l$:
\begin{eqnarray}
  M_{l\parallel }^S+M_{l\perp }^A &=& 1\,, \label{eq:M1} \\
  M_{l\parallel}^A-M_{l\perp}^S &=& 0\,.  \label{eq:M2}
\end{eqnarray} Noticing that $M_{l\parallel}^{\mu}$ and $M_{l\perp}^{\mu}$ only depends on $l$ through its band index, we get four linear equations for the four parameters $\tau_l^S$ and $\tau_l^A$.

Our approach here is a generalization treatments in \cite{SL03,Sinova07} which is only correct for single band situation. In the above two references the authors derived the action of the collision integral $I$ on $|\vec{v}_l|\cos\phi_l$ and $|\vec{v}_l|\sin\phi_l$ as a 2 by 2 matrix and directly inverted it to obtain a solution. This fails when there are multiple bands because the relaxation times calculated this way is in general different for different bands unless there are special symmetries(this was not a problem in \cite{Sinova07} because the two bands in graphene are related by particle-hole symmetry), and that means the operator $I$ doesn't preserve the subspace spanned by $|\vec{v}_l|\cos\phi_l$, $|\vec{v}_l|\sin\phi_l$ and can't be directly inverted.

  Next, we turn to evaluation of the above results. We choose the temperature gradient $\nabla T$ to be along the $\hat{x}$ direction, such that $\phi_l$ coincides with $\theta_l=\theta_k$ defined previously. Plugging in the scattering rates $\gamma_{\alpha\beta}(p,q)$, we obtain
\begin{equation}\label{eq:Mparallel}
  M_{\alpha\parallel}^{\mu}=\frac{E^3 n_i \left(a^2 w_{\alpha }^4 w_{\bar{\alpha }}^2 \left(3 w_{\bar{\alpha }}^4+w_{\alpha }^4\right) \tau _{\alpha }^{\mu }-2 a b E^2 w_{\alpha }^2 w_{\bar{\alpha }}^2 \left(3 w_{\bar{\alpha }}^4 \tau _{\alpha }^{\mu }+w_{\alpha }^4 \tau _{\bar{\alpha }}^{\mu
   }\right)+4 b^2 E^4 \left(w_{\bar{\alpha }}^6+w_{\alpha }^6\right) \tau _{\alpha }^{\mu }\right)}{8 \rho ^2 w_{\alpha }^{10} w_{\bar{\alpha }}^6}\,,
\end{equation}
\begin{equation}\label{eq:Mperp}
  M_{\alpha\perp}^{\mu}=\frac{a b E^8 \eta  n_i \left(w_{\bar{\alpha }}^4 \tau _{\alpha }^{\mu }+w_{\alpha }^4 \tau _{\bar{\alpha }}^{\mu }\right) \left(a w_{\alpha }^2 w_{\bar{\alpha }}^2 \left(w_{\bar{\alpha }}^2+w_{\alpha }^2\right)-2 b E^2 \left(w_{\alpha }^2 w_{\bar{\alpha
   }}^2+w_{\bar{\alpha }}^4+w_{\alpha }^4\right)\right)}{16 \rho ^4 w_{\alpha }^{14} w_{\bar{\alpha }}^{10}}\,.
\end{equation} Here $\alpha=1,2$ is the band index, and $\bar{\alpha}=3-\alpha$. The solution of \eqref{eq:M1} and \eqref{eq:M2} is given in \eqref{eq:taus}.

\section{Numerical Inversion of $I_S$}\label{sec:invertIS}

  The symmetric collision rate $\gamma^S_{ll'}$ is a low-rank matrix. It is therefore numerically beneficial to invert the symmetric collision integral $I_S$ on this reduced subspace. Our treatment follows \cite{Cole2000}.

  As we will show later, at quadratic order the scattering rate is separable, in the form
\begin{equation}\label{}
  \gamma_{ll'}^S=\sum_{a}S_a U^a_l V^a_{l'}\delta(E_l-E_{l'}),
\end{equation} where $S_a$ are constants.

  Inverting $I_S$ is equivalent to solving the following equation
\begin{equation}\label{eq:BEL}
  \sum_{l'}(\gamma^S_{l'l}L^i_l-\gamma^S_{ll'}L^i_{l'})=v_l^i,
\end{equation}where $l,l'$ label states. Here $v_l^i$ on the RHS refers to the $i$-th component of velocity of state $l$, but our result applies to arbitrary function $v_l^i$ of states $l$.

  The ansatz we shall use is
\begin{equation}\label{eq:BELansatz}
  L^i_{l}=v_l^i \tau_l+\tau_l\sum_{a} r^a(E_l) U^a_l,
\end{equation}where
\begin{equation}\label{}
  \frac{1}{\tau_l}=\sum_{l'}\gamma^S_{l'l},
\end{equation} and the undetermined coefficient $r_a$ is a function of energy only.

  Eq.\eqref{eq:BEL} now takes the form
\begin{equation}\label{}
  L_l^i=v_l^i \tau_l+\tau_l \sum_{l'}\sum_{a}S_a U^a_l V^a_{l'}\delta(E_l-E_{l'}) L^i_{l'}.
\end{equation} Comparing with the ansatz \eqref{eq:BELansatz}, we have
\begin{equation}\label{}
  r^a(E_l)=\sum_{b}W^{ab}(E_l) r^b(E_l)+K^a(E_l),
\end{equation} and the matrix $W^{ab}_l$ is given by
\begin{equation}\label{}
  W^{ab}(E_l)=\sum_{l'}\delta(E_{l'}-E_l) S_a V^a_{l'} U^b_{l'}\tau_{l'},
\end{equation} and the inhomogeneous term is
\begin{equation}\label{}
  K^a(E_l)=\sum_{l'}\delta(E_{l'}-E_l) S_a  V^a_{l'} \tau_{l'} v^i_{l'},
\end{equation}
Therefore the coefficients can be obtained as
\begin{equation}\label{}
  r^a(E_l)=\left(\frac{1}{1-W(E_l)}\right)^{ab}K^b(E_l).
\end{equation} This is manageable as for each energy we just need to invert a small (size of hundreds instead of thousands) dimensional matrix.
The matrix $W(E_l)$ contains a unit eigenvector which corresponds to the zero mode $L_l^i=const.$. Therefore the solution is ambiguous by a zero mode, which has no effect on the transport coefficients. Numerically we deal with the zero mode by subtracting the corresponding eigenvectors from $W(E_l)$ such that the eigenvalue becomes zero.

Finally, let's describe how to decompose $\gamma^S_{ll'}$. The starting point is to decompose the $Q$-matrix in \eqref{eq:Q3}:
\begin{equation}
    Q_{JI}(q,p)=\sum_{\bar{a}=1}^{m_Q} \mathcal{F}_{\bar{a}} V^{(Q)}_{\bar{a}J}(q) V^{(Q)}_{\bar{a}I}(p).
\end{equation} Here $\mathcal{F}_{\bar{a}}$ are constants independent of $p,q$. In practice we found a decomposition with $m_Q=12$. To proceed, we follow the calculations in Secs.~\ref{sec:model},\ref{sec:scatter} to compute the amplitude $F_{ll'}$ and scattering rate $\gamma_{ll'}$. We found the $P$-matrix as defined in \eqref{eq:Pmat} to be
\begin{equation}\label{}
  P_{BA}(q,p)=\sum_{\bar{a}=1}^{m_Q} \mathcal{F}_{\bar{a}}  U^{(P)}_{\bar{a}B}(q) V^{(P)}_{\bar{a}A}(p),
\end{equation}where
\begin{equation}\label{}
  U_{\bar{a} B}^{(P)}(q)=\sum_K V^{(Q)}(q)_{\bar{a}K}(M^{-1}(q)^*)^K_{~B}\,, \quad V_{\bar{a} A}^{(P)}(p)=\sum_K V^{(Q)}(p)_{\bar{a}K}M^{-1}(p)^K_{~A}\,.
\end{equation} Here $~*$ means complex conjugation.
When obtaining the $F$-matrix as defined in \eqref{eq:Fdef}, we need to double the rank of the matrix
\begin{equation}\label{}
  F_{\beta\alpha}(q,p)=\sum_{\bar{a}=1}^{2m_Q}\tilde{\mathcal{F}}_{\bar{a}}U_{\bar{a}\beta}^{(F)}(q)V_{\bar{a}\alpha}^{(F)}(p).
\end{equation} Here the new set of basis is obtained by projecting $U^{(P)}$  and $V^{(P)}$ down to the first half and last half entries as in  \eqref{eq:chiA} and \eqref{eq:Fdef}, and $\tilde{\mathcal{F}}$ is a double copy of $\mathcal{F}$. Finally, the scattering rate $\gamma^S_{ll'}$ is the square of the amplitude $F_{ll'}$, therefore the rank also get squared:
\begin{equation}\label{}
  \gamma_{ll'}^S=\sum_{\bar{a},\bar{b}=1}^{2m_Q}S_{\bar{a}\bar{b}} U^{\bar{a}\bar{b}}_l V^{\bar{a}\bar{b}}_{l'}\delta(E_l-E_{l'}),
\end{equation} where
\begin{equation}\label{}
  S_{\bar{a}{\bar{b}}}=\frac{2\pi n_i}{V_s} \tilde{\mathcal{F}}_{\bar{a}}\tilde{\mathcal{F}}_{\bar{b}}^*\,,
\end{equation}
\begin{equation}\label{}
  U_l^{\bar{a}\bar{b}}=U_{\bar{a}\alpha}^{(F)}(p)U_{\bar{b}\alpha}^{(F)}(p)^*\,,\quad l=(p\alpha)\,,
\end{equation}
\begin{equation}\label{}
  V_l^{\bar{a}\bar{b}}=V_{\bar{a}\alpha}^{(F)}(p)V_{\bar{b}\alpha}^{(F)}(p)^*\,,\quad l=(p\alpha)\,.
\end{equation} This is the desired decomposition.

\section{Explicit values of $\hat{\Gamma}$, $\hat{\gamma}^S$, $\hat{\gamma}^A$}\label{sec:gammaval}

\begin{equation}\label{}
 \hat{\Gamma}_\alpha=
\left(
\begin{array}{c}
 \frac{E^3 n_i \left(a^2 w_2^2 w_1^8+3 a^2 w_2^6 w_1^4+4 b^2 E^4 w_1^6+4 b^2 E^4 w_2^6\right)}{8 \rho ^2 w_1^{10} w_2^6} \\
 \frac{E^3 n_i \left(a^2 w_1^2 w_2^8+3 a^2 w_1^6 w_2^4+4 b^2 E^4 w_2^6+4 b^2 E^4 w_1^6\right)}{8 \rho ^2 w_1^6 w_2^{10}} \\
\end{array}
\right)\,.
\end{equation}
\begin{eqnarray}
\hat{\gamma }_{ 0}^S&=&\left(
\begin{array}{cc}
 \frac{E^3 n_i \left(3 a^2 w_1^4+4 b^2 E^4\right)}{8 \rho ^2 w_1^{10}} & \frac{E^3 n_i \left(a^2 w_1^2 w_2^2+4 b^2 E^4\right)}{8 \rho ^2 w_1^4 w_2^6} \\
 \frac{E^3 n_i \left(a^2 w_1^2 w_2^2+4 b^2 E^4\right)}{8 \rho ^2 w_1^6 w_2^4} & \frac{E^3 n_i \left(3 a^2 w_2^4+4 b^2 E^4\right)}{8 \rho ^2 w_2^{10}} \\
\end{array}
\right)\,,\\
\hat{\gamma }_{\pm 1}^S&=&\left(
\begin{array}{cc}
 \frac{3 a b E^5 n_i}{4 \rho ^2 w_1^8} & \frac{a b E^5 n_i}{4 \rho ^2 w_1^3 w_2^5} \\
 \frac{a b E^5 n_i}{4 \rho ^2 w_1^5 w_2^3} & \frac{3 a b E^5 n_i}{4 \rho ^2 w_2^8} \\
\end{array}
\right)\,,\\
\hat{\gamma }_{\pm 2}^S&=&\left(
\begin{array}{cc}
 \frac{E^3 n_i \left(a^2 w_1^4+b^2 E^4\right)}{4 \rho ^2 w_1^{10}} & -\frac{b^2 E^7 n_i}{4 \rho ^2 w_1^4 w_2^6} \\
 -\frac{b^2 E^7 n_i}{4 \rho ^2 w_1^6 w_2^4} & \frac{E^3 n_i \left(a^2 w_2^4+b^2 E^4\right)}{4 \rho ^2 w_2^{10}} \\
\end{array}
\right)\,,\\
\hat{\gamma }_{\pm 3}^S&=&\left(
\begin{array}{cc}
 \frac{a b E^5 n_i}{4 \rho ^2 w_1^8} & -\frac{a b E^5 n_i}{4 \rho ^2 w_1^3 w_2^5} \\
 -\frac{a b E^5 n_i}{4 \rho ^2 w_1^5 w_2^3} & \frac{a b E^5 n_i}{4 \rho ^2 w_2^8} \\
\end{array}
\right)\,,\\
\hat{\gamma }_{\pm 4}^S&=&\left(
\begin{array}{cc}
 \frac{a^2 E^3 n_i}{16 \rho ^2 w_1^6} & -\frac{a^2 E^3 n_i}{16 \rho ^2 w_1^2 w_2^4} \\
 -\frac{a^2 E^3 n_i}{16 \rho ^2 w_1^4 w_2^2} & \frac{a^2 E^3 n_i}{16 \rho ^2 w_2^6} \\
\end{array}
\right)\,.
\end{eqnarray}
\begin{eqnarray}\label{eq:hatA1}
\hat{\gamma }_{\pm 1}^A&=&\pm \frac{a b E^8 \eta  n_i \left(-2 b E^2 w_1^4-2 b E^2 w_1^2 w_2^2+a w_1^4 w_2^2-2 b E^2 w_2^4+a w_1^2 w_2^4\right)}{16 \rho ^4 w_1^{14} w_2^{14}} \left(
\begin{array}{cc}
 -i w_2^8 & -i w_1^5 w_2^3 \\
 -i w_1^3 w_2^5 & -i w_1^8 \\
\end{array}
\right)\,,\\
\hat{\gamma }_{\pm 2}^A&=&\pm \frac{b^3 E^{12} \eta  n_i \left(w_1^2-w_1 w_2+w_2^2\right) \left(w_1^2+w_1 w_2+w_2^2\right)}{4 \rho ^4 w_1^{16} w_2^{16}} \left(
\begin{array}{cc}
 i w_2^{10} & i w_1^6 w_2^4 \\
 i w_1^4 w_2^6 & i w_1^{10} \\
\end{array}
\right)\,,\\
\hat{\gamma }_{\pm 3}^A&=&\pm \frac{a b E^8 \eta  n_i \left(2 b E^2 w_1^4+2 b E^2 w_1^2 w_2^2+a w_1^4 w_2^2+2 b E^2 w_2^4+a w_1^2 w_2^4\right)}{16 \rho ^4 w_1^{14} w_2^{14}} \left(
\begin{array}{cc}
 i w_2^8 & i w_1^5 w_2^3 \\
 i w_1^3 w_2^5 & i w_1^8 \\
\end{array}
\right)\,,\\
\hat{\gamma }_{\pm 4}^A&=&\pm \frac{a^3 E^6 \eta  n_i \left(w_1^2+w_2^2\right)}{32 \rho ^4 w_1^{10} w_2^{10}} \left(
\begin{array}{cc}
 i w_2^6 & i w_1^4 w_2^2 \\
 i w_1^2 w_2^4 & i w_1^6 \\
\end{array}
\right)\,.\label{eq:hatA4}
\end{eqnarray}

\section{Hydrodynamic corrections to the Boltzmann equation with boundary}
\label{sec:BoltzBoundary}
In this part we present a detailed analysis of Eq.~\eqref{eq:BEphi1}.

  From Eqs.\eqref{eq:hatA1}-\eqref{eq:hatA4}, $\hat{\gamma}^A$ only acts on first through fourth harmonics and has opposite signs for positive and negative harmonics, so $\hat{\gamma}^A$ actually transforms $\cos n\theta$ into $\sin n\theta$ since
\begin{equation}\label{}
  \hat{\gamma}^A[w_\alpha\cos n\theta]_\alpha=\hat{\gamma}^A[w_\alpha\frac{e^{in\theta}+e^{-in\theta}}{2}]_\alpha=i\sin n\theta\sum_\beta\hat{\gamma}^A_{+n,\alpha\beta}w_\beta\,.
\end{equation}

  We therefore need to calculate the harmonic decomposition of $\phi^{(0)}$. From the symmetry $\phi^{(0)}(\theta)=-\phi^{(0)}(\pi-\theta)$ and $\phi^{(0)}(-\theta)=-\phi^{(0)}(-\pi+\theta)$ for $0<\theta<\pi$ , we see $\phi^{(0)}$ can only contain $\cos \theta,\cos 3\theta$ and $\sin2\theta,\sin 4\theta$ harmonics. The $\cos\theta$ and $\cos3\theta$ harmonics are enhanced by a logarithmic factor, whose coefficients are
\begin{equation}\label{}
  \phi^{(0)}_{\alpha,c1}=\int_{-\pi}^{\pi}\frac{\rd \theta}{\pi}\phi^{(0)}_\alpha(\theta)\cos\theta=\frac{2}{\pi}X\left(y\ln\frac{w_\alpha}{\hat{\Gamma}_\alpha y}+(W-y)\ln\frac{w_\alpha}{\hat{\Gamma}_\alpha(W-y)}\right)=\frac{2}{\pi}X W\ln\frac{w_\alpha}{\hat{\Gamma}_\alpha W}\,,
\end{equation}
\begin{equation}\label{}
  \phi^{(0)}_{\alpha,c3}=\int_{-\pi}^{\pi}\frac{\rd \theta}{\pi}\phi^{(0)}_\alpha(\theta)\cos3\theta=\frac{2}{\pi}X\left(y\ln\frac{w_\alpha}{\hat{\Gamma}_\alpha y}+(W-y)\ln\frac{w_\alpha}{\hat{\Gamma}_\alpha(W-y)}\right)=\frac{2}{\pi}X W\ln\frac{w_\alpha}{\hat{\Gamma}_\alpha W}\,,
\end{equation}where we have only retained the leading term in $\hat{\Gamma}_\alpha W/w_\alpha$. The $\sin2\theta$ and $\sin 4\theta$ harmonics are subdominant by a log factor, but we should still keep them because $\hat{\gamma}^A_{2,4}$ are parametrically larger than $\hat{\gamma}^A_{1,3}$.
\begin{equation}\label{}
  \phi_{\alpha,s2}^{(0)}=\int_{-\pi}^\pi\frac{\rd \theta}{\pi}\phi_{\alpha}^{(0)}\sin 2\theta=X(2y-W) \,,
\end{equation}
\begin{equation}\label{}
  \phi_{\alpha,s4}^{(0)}=\int_{-\pi}^\pi\frac{\rd \theta}{\pi}\phi_{\alpha}^{(0)}\sin 4\theta=X(2y-W)\,.
\end{equation} Here we have also just kept the leading order term in $\hat{\Gamma}_\alpha W/w_\alpha$.

Therefore the RHS of \eqref{eq:BEphi1} is
\begin{equation}\label{eq:BEphi1RHS}
  \hat{\gamma}^A[\phi^{(0)}]_\alpha(\theta)=\frac{2}{\pi}X W(z_{s1,\alpha}\sin\theta+z_{s3,\alpha}\sin3\theta)+X(2y-W)(z_{c2,\alpha}\cos2\theta+z_{c4,\alpha}\cos 4\theta)\,,
\end{equation}where
\begin{equation}\label{}
\begin{split}
  z_{s1,\alpha}&=\sum_{\beta}i\hat{\gamma}^{A}_{+1,\alpha\beta}\ln\frac{w_\beta}{\hat{\Gamma}_\beta W}\,,\\
  z_{s3,\alpha}&=\sum_\beta i\hat{\gamma}^{A}_{+3,\alpha\beta}\ln\frac{w_\beta}{\hat{\Gamma}_\beta W} \,, \\
  z_{c2,\alpha}&=\sum_\beta -i\hat{\gamma}^A_{+2,\alpha\beta}\,, \\
  z_{c4,\alpha}&=\sum_\beta -i\hat{\gamma}^A_{+4,\alpha\beta}\,.
\end{split}
\end{equation}

The solution of Eq. \eqref{eq:BEphi1} doesn't satisfy zero boundary condition, so we should be more careful about it. The generic solution takes the form
\begin{equation}\label{}
  \phi^{(1)}_{\alpha}(y,\theta)=\phi^{i}_{\alpha}(y,\theta)+C_\alpha(\theta)e^{-\frac{\hat{\Gamma}_\alpha y}{w_\alpha \sin\theta}},
\end{equation} where $\phi^{i}_{\alpha}(\theta)$ is an inhomogeneous solution satisfying the RHS but not the boundary conditions, and the second term is a homogeneous solution to be determined from boundary conditions. We shall use $C_\alpha^{\pm}(\theta)$ to denote the branches of $0<\theta<\pi$ and $-\pi<\theta<0$ respectively. From the boundary conditions \eqref{eq:BC} we can determine
\begin{eqnarray}
  C_{\alpha}^+(\theta) &=& c_D-\phi^{D}_\alpha(\theta) \,,\\
  C_{\alpha}^{-} (\theta)&=& e^{\frac{W \hat{\Gamma}_\alpha}{w_\alpha \sin\theta}}(c_U-\phi^{U}_\alpha(\theta))\,,
\end{eqnarray} where
$\phi_{\alpha}^D(\theta)=\phi^{i}(y=0,\theta),~\phi_{\alpha}^U(\theta)=\phi^i(y=W,\theta)$. Using \eqref{eq:cDcU}, we have
\begin{equation}\label{}
\begin{split}
  c_D&=-\frac{1}{4}\sum_\alpha\int_{-\pi}^0 \rd \theta \sin\theta \left[\phi^D_\alpha(\theta)+C_{\alpha}^-(\theta)\right]\\
&=\frac{1}{4}\sum_{\alpha}\int_0^\pi \rd \theta\sin\theta\left[\phi_\alpha^D(-\theta)-e^{-\frac{W \hat{\Gamma}_\alpha}{w_\alpha \sin\theta}}\phi_\alpha^U(-\theta)+e^{-\frac{W \hat{\Gamma}_\alpha}{w_\alpha \sin\theta}}c_U\right]\,,
\end{split}
\end{equation}
\begin{equation}\label{}
  \begin{split}
     c_U & =\frac{1}{4}\sum_{\alpha}\int_0^\pi  \rd \theta\sin\theta\left[\phi_\alpha^U(\theta)+C_\alpha^+(\theta)e^{-\frac{W \hat{\Gamma}_\alpha}{w_\alpha \sin\theta}}\right] \\
       & =\frac{1}{4}\sum_{\alpha}\int_0^\pi \rd \theta\sin\theta\left[\phi_\alpha^U(\theta)-e^{-\frac{W \hat{\Gamma}_\alpha}{w_\alpha \sin\theta}} \phi_\alpha ^D(\theta)+e^{-\frac{W \hat{\Gamma}_\alpha}{w_\alpha \sin\theta}} c_D\right]\,.
  \end{split}
\end{equation} This yields the solution for $c_D,c_U$ as
  \begin{eqnarray}
    c_D &=& \frac{f_D+f_U-\xi f_U}{\xi(2-\xi)}\,,\\
    c_U &=& \frac{f_D+f_U-\xi f_D}{\xi(2-\xi)}\,,
  \end{eqnarray}
where
\begin{equation}\label{}
  \begin{split}
     \xi &=  1-\frac{1}{4}\sum_{\alpha}\int _0^\pi \rd \theta \sin\theta e^{-\frac{W \hat{\Gamma}_\alpha}{w_\alpha \sin\theta}}\,, \\
     f_D &=   \frac{1}{4}\sum_{\alpha}\int_0^\pi \rd \theta\sin\theta\left[\phi_\alpha^D(-\theta)-e^{-\frac{W \hat{\Gamma}_\alpha}{w_\alpha \sin\theta}}\phi_\alpha^U(-\theta)\right]\,, \\
     f_U &=   \frac{1}{4}\sum_{\alpha}\int_0^\pi \rd \theta\sin\theta\left[\phi_\alpha^U(\theta)-e^{-\frac{W \hat{\Gamma}_\alpha}{w_\alpha \sin\theta}} \phi_\alpha ^D(\theta)\right]\,.
  \end{split}
\end{equation} Here $\xi$ is of smallness $W \hat{\Gamma}_\alpha/w_\alpha$.

For the four inhomogeneous terms in \eqref{eq:BEphi1RHS}, the corresponding inhomogeneous solutions are
\begin{eqnarray}
  \phi^i_{s1,\alpha}&=&\frac{2 X W }{\pi \hat{\Gamma}_\alpha}z_{s1,\alpha}\sin\theta \,,\\
  \phi^i_{s3,\alpha}&=&\frac{2 X W }{\pi \hat{\Gamma}_\alpha}z_{s3,\alpha}\sin3\theta \,,\\
  \phi^i_{c2,\alpha} &=& \frac{X\left((2y-W)\hat{\Gamma}_\alpha-2w_\alpha\sin\theta\right)}{\hat{\Gamma}_\alpha^2}z_{c2,\alpha}\cos2\theta\,,\\
\phi^i_{c4,\alpha} &=&\frac{X\left((2y-W)\hat{\Gamma}_\alpha-2w_\alpha\sin\theta\right)}{\hat{\Gamma}_\alpha^2}z_{c4,\alpha}\cos4\theta\,.
\end{eqnarray}
Notice that these solutions satisfy $\phi_\alpha^D(\theta)=-\phi_\alpha^U(-\theta)$. The corresponding $f_D$ and $f_U$ are
\begin{equation}\label{}
  f_{U,s1,\alpha}=-f_{D,s1,\alpha}=\frac{X W^2}{\pi }\sum_{\alpha} \frac{z_{s1,\alpha} }{w_\alpha}\,,
\end{equation}
\begin{equation}\label{}
  f_{U,s3,\alpha}=-f_{D,s3,\alpha}=\frac{X W^2}{3\pi }\sum_{\alpha} \frac{z_{s3,\alpha} }{w_\alpha}\,,
\end{equation}
\begin{equation}\label{}
  f_{U,c2,\alpha}=-f_{D,c2,\alpha}\sim\frac{X W^2}{12}\sum_{\alpha}\frac{ W \hat{\Gamma}_\alpha}{ w_\alpha^2}z_{c2,\alpha}\ln\frac{w_\alpha}{W\hat{\Gamma}_\alpha}\,,
\end{equation}
\begin{equation}\label{}
  f_{U,c4,\alpha}=-f_{D,c4,\alpha}\sim\frac{X W^2}{12}\sum_{\alpha}\frac{ W \hat{\Gamma}_\alpha}{ w_\alpha^2}z_{c4,\alpha}\ln\frac{w_\alpha}{W\hat{\Gamma}_\alpha}\,.
\end{equation} Here we have calculated to the lowest non-trivial order in $\hat{\Gamma}_\alpha$. Since $\xi$ is also proportional to $\hat{\Gamma}_\alpha$, we get $c_U=-c_D=\frac{1}{2}f_U$. The leading order terms in $f_{c2}$ and $f_{c4}$ cancelled and so the remaining terms only serve as scale estimates and the numerical coefficient can't be trusted. This completes the solution of \eqref{eq:BEphi1}.

As discussed in the main text, to extract the thermal hall conductivity we need to calculate the difference of zeroth harmonics of $\phi^{(1)}_{\alpha}$ between $y=W$ and $y=0$, which is given by
\begin{equation}\label{}
  \Delta\phi_{\alpha,m=0}\vert_{y=0}^{y=W}=\int_0^\pi \frac{\rd \theta}{\pi} [c_U-\phi_\alpha^U(-\theta)](1-e^{-\frac{\hat{\Gamma}_\alpha W}{w_\alpha\sin\theta}})\,.
\end{equation} The explicit contributions for the four terms are
\begin{equation}\label{}
  \Delta \phi_{s1,\alpha,m=0}\vert_{y=0}^{y=W}=\frac{2 X W^2}{\pi}\frac{z_{s1,\alpha}}{w_\alpha}\,,
\end{equation}
\begin{equation}\label{}
  \Delta \phi_{s3,\alpha,m=0}\vert_{y=0}^{y=W}=\frac{2 X W^2}{\pi}\frac{z_{s3,\alpha}}{w_\alpha}\,,
\end{equation}
\begin{equation}\label{}
  \Delta \phi_{c2,\alpha,m=0}\vert_{y=0}^{y=W}\sim \Delta \phi_{c4,\alpha,m=0}\vert_{y=0}^{y=W}\sim X W^2 \frac{z_{c2/c4,\alpha}}{w_\alpha} \ln \frac{w_\alpha}{\hat{\Gamma}_\alpha W}\,.
\end{equation} As before the last two integrals are only scale estimates.

\bibliography{phonon}
\end{document}